\documentclass[aps,prd,twocolumn,showpacs,superscriptaddress,floatfix
]{revtex4-2}

\usepackage{amsmath}
\usepackage{graphicx}
\usepackage{dcolumn}
\usepackage{bm}
\usepackage{hyperref}
\usepackage[utf8]{inputenc}
\usepackage{multirow}
\usepackage{soul}
\usepackage{float}
\usepackage{graphicx}
\usepackage{times}
\usepackage[normalem]{ulem}
\usepackage{color}
\usepackage{cellspace}
\usepackage{lipsum, babel}
\usepackage{lipsum}
\usepackage{xcolor}
\definecolor{customviolet}{RGB}{148,0,211}
\newcommand{\be}{\begin{equation}}
\newcommand{\ee}{\end{equation}}
\newcommand{\bea}{\begin{eqnarray}}
\newcommand{\eea}{\end{eqnarray}}

\newcommand{\comment}[1]{}
\renewcommand\sout{\bgroup \color{red} \ULdepth=-.5ex \ULset}
\def\simge{\mathrel{\rlap{\raise 0.511ex
     \hbox{$>$}}{\lower 0.511ex \hbox{$\sim$}}}}
\def\simle{\mathrel{\rlap{\raise 0.511ex
      \hbox{$<$}}{\lower 0.511ex \hbox{$\sim$}}}}

\usepackage{multirow}

\DeclareUnicodeCharacter{2212}{-}

\begin{document}


\title{Exploring robust correlations between fermionic dark matter model parameters and neutron star properties: A two-fluid perspective} 

\author{Prashant Thakur}
\email{p20190072@goa.bits-pilani.ac.in}
\affiliation{Department of Physics, BITS-Pilani, K. K. Birla Goa Campus, Goa 403726, India}
\author{Tuhin Malik}
\email{tuhin.malik@uc.pt}
\affiliation{CFisUC, Department of Physics, University of Coimbra, P-3004 - 516  Coimbra, Portugal}
\author{Arpan Das}
\email{arpan.das@ifj.edu.pl}
\affiliation{Institute  of  Nuclear  Physics  Polish  Academy  of  Sciences,  PL-31-342  Krak\'ow,  Poland}
\author{T. K. Jha}
\email{tkjha@goa.bits-pilani.ac.in}
\affiliation{Department of Physics, BITS-Pilani, K. K. Birla Goa Campus, Goa 403726, India}
\author{Constança Providência}
\email{cp@uc.pt}
\affiliation{CFisUC, 
	Department of Physics, University of Coimbra, P-3004 - 516  Coimbra, Portugal}

\date{\today}

\begin{abstract} 
We investigate the probable existence of dark matter in the interior of neutron stars.
Despite the current state of knowledge, the observational properties of neutron stars have not definitively ruled out the possibility of dark matter. Our research endeavors to shed light on this intriguing mystery by examining how certain neutron star properties, including mass, radius, and tidal deformability, might serve as constraints for the dark matter model.

In our investigation, we adopt a two-fluid approach to calculate the properties of neutron stars. For the nuclear matter EOS, we employ several realistic EOS derived from the relativistic mean field model (RMF), each exhibiting varying stiffness and composition. In parallel, we look into the dark matter EOS, considering fermionic matter with repulsive interaction described by a relativistic mean field Lagrangian. A reasonable range of parameters is sampled meticulously.

Our study primarily focuses on exploring correlations between the dark matter model parameters and different neutron star properties using a rich set of EOSs. Interestingly, our results reveal a promising correlation between the dark matter model parameters and stellar properties, particularly when we ignore the uncertainties in the nuclear matter EOS. However, when introducing uncertainties in the nuclear sector, the correlation weakens, suggesting that the task of conclusively constraining any particular dark matter model might be challenging using global properties alone, such as mass, radius, and tidal deformability.

Notably, we find that dark-matter admixed stars tend to have higher central baryonic density, potentially allowing for non-nucleonic degrees of freedom or direct Urca processes in stars with lower masses. There is also a tantalizing hint regarding the detection of stars with the same mass but different surface temperatures, which may indicate the presence of dark matter. With our robust and extensive dataset, we delve deeper and demonstrate that even in the presence of dark matter, the semi-universal C-Love relation remains intact. This captivating finding adds another layer of complexity to the interplay between dark matter and neutron star properties.
\end{abstract}

\maketitle
\section{Introduction} \label{sec1}
Neutron stars (NSs), with their compact nature, hold great fascination in the vast expanse of the cosmos. Although they are of great significance in astrophysics, the interiors of neutron stars remain mysterious \cite{1996cost.book.....G, book.Haensel2007, Rezzolla:2018jee}. One of the most intriguing puzzles is the nature of the dense matter that forms their cores. It is believed that these cores can reach densities 5-10 times greater than normal nuclear saturation density. However, the exact composition and behavior of this extreme matter under such extreme conditions elude our understanding. To unravel the secrets concealed within these dense cores, scientists have devoted considerable effort to studying the equation-of-state (EOS), which characterizes the interplay among pressure, density, and temperature within a specific substance~\cite{Lattimer:2000nx}. 

The issue of galaxy rotation curves stands out as a prominent indication that galaxies may not be only composed of ordinary nuclear matter~\cite{Rubin:1978kmz}. These rotation curves, which depict the rotational velocities of stars and gas in galaxies, exhibit unexpected behavior that cannot be explained solely by the presence of visible matter. This suggests the existence of an additional component known as dark matter~\cite{Bauer:2017qwy}. Dark matter, as its name suggests, does not interact directly with electromagnetic radiation and remains invisible to traditional telescopes. Its existence can be inferred indirectly through its gravitational effects on visible matter. The dense cores within neutron stars can act as gravitational traps for dark matter particles, potentially influencing their behavior and properties.
The highly dense matter inside neutron stars \cite{Bertone:2007ae,deLavallaz:2010wp,Guver:2012ba} can enhance the dark matter capture process inside these objects. Several theoretical models propose the presence of dark matter within neutron star cores, with one possibility being the accumulation of dark matter particles, such as Weakly Interacting Massive Particles (WIMPs), due to gravitational attraction \cite{Goldman:1989nd}. Significant accumulation of non-self annihilating dark matter inside the
compact objects can affect the structure of these compact objects~\cite{Raj:2017wrv,Goldman:1989nd,Gould:1989gw,Kouvaris:2007ay,Kouvaris:2010vv,deLavallaz:2010wp,Guver:2012ba,Ellis:2018bkr,Panotopoulos:2017idn,Das:2018frc,Das:2020vng}. From a particle physics point of view so far there are many candidates for dark matter particles, such as bosonic dark matter, axions, sterile neutrinos, and different possible WIMPs
have been proposed in literature~\cite{Bertone:2010zza,Bauer:2017qwy,Calmet:2020pub}. Since the nature of particle dark matter is uncertain, both bosonic and fermionic dark matter particles have been considered to study its effect on the neutron star dynamics~\cite{Ellis:2018bkr,Panotopoulos:2017idn,Diedrichs:2023trk,Leung:2022wcf}. Naively for fermionic dark matter particles modeled by the ideal Fermi degenerate gas, it is the degeneracy pressure that allows for a stable neutron star configuration~\cite{Leung:2022wcf}. But for the bosonic dark matter particle one must include self-interaction to obtain stable neutron star configurations~\cite{Ellis:2018bkr}.  Practically neutron star EOS in the presence of dark matter can be much more complicated.  

The advent of multi-messenger observations of compact astrophysical objects and the discovery of gravitational waves from neutron star-neutron star mergers by advanced LIGO and VIRGO collaborations \cite{abbott2016observation,abbott2017gw170817} have not only opened new avenues for exploring the EOS of dense matter but also offer a fresh perspective on the study of dark matter in compact objects, whether fermionic or bosonic~\cite{Baiotti:2016qnr,Baiotti:2019sew}. Recent advancements in this field of research have shed new light on the presence of dark matter inside neutron stars \cite{Karkevandi:2021ygv,Das:2021yny,Miao:2022rqj,Hippert:2022snq,Rutherford:2022xeb,Das:2018frc}. Progress in numerical simulations and theoretical models has provided insights into the behavior of dark matter within the cores of neutron stars.
Recently, a number of studies have been conducted using statistical Bayesian methods to constrain the EOS of neutron stars using astrophysical observations \cite{coughlin2019multimessenger, dietrich2020multimessenger, o2020parametrized, pang2021nuclear, Pradhan:2022vdf}. In constraining the EOS, the correlation has proven useful, with nuclear matter parameters (NMPs) as its key components \cite{vidana2009density, ducoin2010nuclear}. Modern studies have documented correlations between empirical nuclear parameters and neutron star observables \cite{carson2019constraining, xie2019bayesian, zimmerman2020measuring}. Although some studies have suggested the existence of dark matter (DM) inside neutron stars \cite{Das:2020ecp, Xiang:2013xwa, Kouvaris:2010vv, kouvaris2012limits, bell2019capture, hook2018probing}, up to our knowledge most of these studies have not extensively examined correlations with measurable properties, including dark matter parameters and neutron star astrophysical observable.

The motivation behind this study is to explore potential correlations between the dark matter sector and the global properties of neutron stars while taking into account the inherent uncertainties in the EOS within the baryonic sector. It is crucial to consider the uncertainties associated with the EOS in the baryonic sector, as they can affect our understanding of the overall behavior of neutron stars.
 
Using the well-known Kendall rank correlation studies we look into the relationship between the dark matter sector and neutron star properties. We seek to uncover any connections or influences that may exist, providing valuable insights into both the properties of dark matter and the behavior of these enigmatic cosmic objects.  Furthermore, we aim to explore new avenues to constrain the dark matter sector by investigating its effects on neutron star cooling, which is a key observable, or by examining the viability of direct Urca processes. 

The paper has the following organization. In Section ~\ref{sec2}, we introduce the basic formalism of the equation-of-state for nuclear matter and dark matter, as well as the two-fluid formalism of the Tolman-Oppenheimer-Volkoff equation and the Kendall rank correlation coefficient. In Section ~\ref{results}, we present and discuss the results of the current study. Finally, in Section ~\ref{sec5}, we provide concluding remarks.

\section{Methodology}\label{sec2}

\subsection{Nuclear matter EOS}\label{SubsecB1}
Our study considers the relativistic mean field (RMF) description \cite{Malik:2023mnx} of the nuclear matter EOS: a mean-field theory approach that includes nonlinear meson terms, both self-interactions and mixed terms.

\begin{equation}
  \mathcal{L}=   \mathcal{L}_N+ \mathcal{L}_M+ \mathcal{L}_{NL}
\end{equation} 
with
\begin{equation}
\begin{aligned}
\mathcal{L}_{N}=& \bar{\Psi}\Big[\gamma^{\mu}\left(i \partial_{\mu}-g_{\omega} \omega_{\mu}-
g_{\varrho} {\boldsymbol{t}} \cdot \boldsymbol{\varrho}_{\mu}\right) \\
&-\left(m-g_{\sigma} \sigma\right)\Big] \Psi \\
\mathcal{L}_{M}=& \frac{1}{2}\left[\partial_{\mu} \sigma \partial^{\mu} \sigma-m_{\sigma}^{2} \sigma^{2} \right] \\
&-\frac{1}{4} F_{\mu \nu}^{(\omega)} F^{(\omega) \mu \nu} 
+\frac{1}{2}m_{\omega}^{2} \omega_{\mu} \omega^{\mu} \nonumber\\
&-\frac{1}{4} \boldsymbol{F}_{\mu \nu}^{(\varrho)} \cdot \boldsymbol{F}^{(\varrho) \mu \nu} 
+ \frac{1}{2} m_{\varrho}^{2} \boldsymbol{\varrho}_{\mu} \cdot \boldsymbol{\varrho}^{\mu}.\\
    			\mathcal{L}_{NL}=&-\frac{1}{3} b ~m ~g_\sigma^3 \sigma^{3}-\frac{1}{4} c g_\sigma^4 \sigma^{4}+\frac{\xi}{4!} g_{\omega}^4 (\omega_{\mu}\omega^{\mu})^{2} \nonumber\\&+\Lambda_{\omega}g_{\varrho}^{2}\boldsymbol{\varrho}_{\mu} \cdot \boldsymbol{\varrho}^{\mu} g_{\omega}^{2}\omega_{\mu}\omega^{\mu},
\end{aligned}
\label{lagrangian}
\end{equation}

\begin{figure*}
 \includegraphics[width=1.0\linewidth]{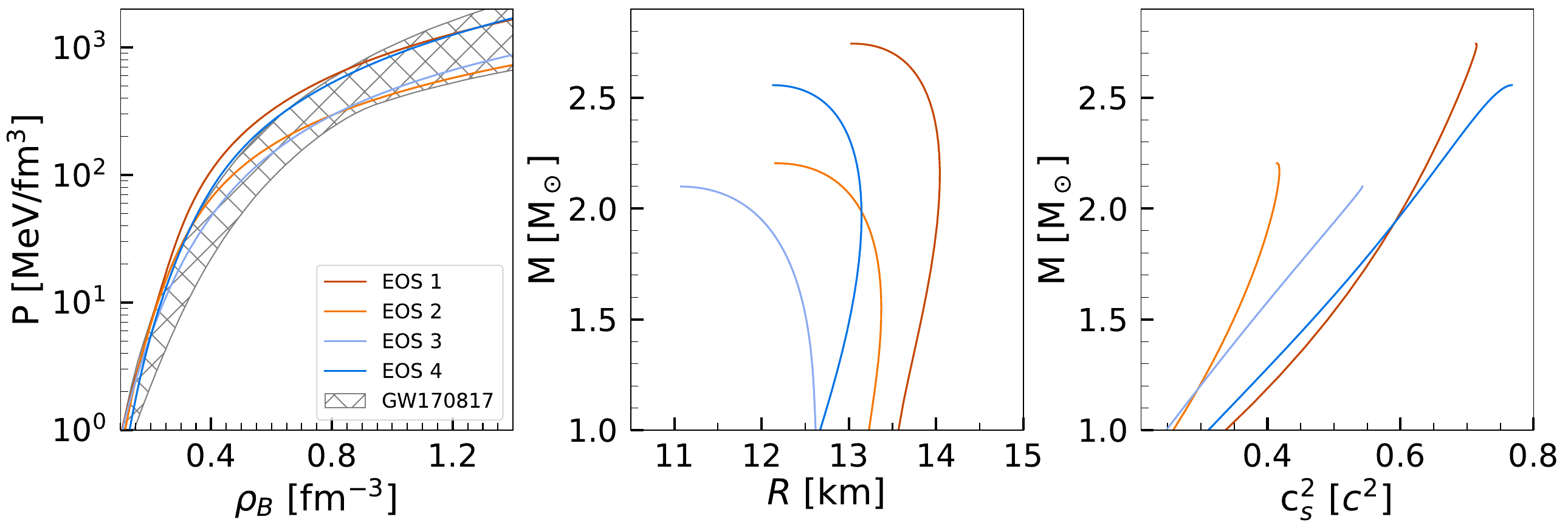}
 \caption{(left plot) The pressure $P$ as a function of baryon density $\rho_B$, (middle plot) the NS mass $M$ as a function of radius $R$, and (right plot) NS mass $M$ as a function of the square of the speed of sound $c_s^2$ for nuclear matter EOS: EOS1, EOS2, EOS3, and EOS4, respectively.}
 \label{fig:eos}
 \end{figure*}

In this context, the field $\Psi$ represents a Dirac spinor that describes the nucleon doublet (consisting of neutron and proton) with a bare mass $m$. The couplings of the nucleons to the meson fields $\sigma$, $\omega$, and $\varrho$ are denoted by $g_{\sigma}$, $g_{\omega}$, and $g_{\varrho}$ respectively, with corresponding masses $m_\sigma$, $m_\omega$, and $m_\varrho$. The parameters $b$, $c$, $\xi$, and $\Lambda_{\omega}$, which determine the strength of the non-linear terms, are determined alongside the couplings $g_i$ (where $i=\sigma, \omega, \varrho$) by imposing a set of constraints.

 We chose four samples from the EOS generated in \cite{Malik:2023mnx}, which we designate as EOS1, EOS2, EOS3, and EOS4. They have been constrained to several nuclear matter properties, in particular, the saturation density, binding energy, incompressibility, symmetry energy at saturation, and the pure neutron matter pressure calculated with chiral effective field theory. It was also imposed that the pure neutron matter pressure must be an increasing function of the baryonic density and stars with at least 2.0 $M_{\odot}$ must be described. The parameters of all these four models are presented in Table \ref{tableI}.  Figure \ref{fig:eos} displays these four EOS, along with their corresponding neutron stars (NS) properties. The graph depicts the pressure, denoted as $P$, as a function of baryon density $\rho_B$ in the left plot. The middle plot showcases the NS mass, denoted as $M$, as a function of radius $R$. Additionally, the right plot illustrates the relation between the NS mass $(M)$ and the square of the speed of sound (c$_s^2$) for the four nuclear matter EOSs. The EOS1 is the stiffest and EOS3 is the softest one.  The nuclear saturation properties along with star properties can be accessed from Table \ref{tableII}. It can be seen from the table that the NS maximum mass of these four EOSs ranges from 2.10 to 2.74 $M_{\odot}$. The radius and tidal deformability for a 1.4 $M_{\odot}$ NS are in the range of 12.55--13.78 km and 462--844, respectively.

\begin{table*}[]
\setlength{\tabcolsep}{4.pt}
      \renewcommand{\arraystretch}{1.1}
\begin{tabular}{cccccccc}
\toprule
EOS &         $g_\sigma$ &         $g_\omega$ &         $g_\rho$ &         $B$ &         $C$ &        $\xi$ &       $\Lambda_\omega$ \\
\hline
EOS1   &  10.411847 &  13.219028 &  11.180337 &  2.541001 & -3.586261 &  0.000845 &  0.027999 \\
EOS2 &   11.150279 &  14.420375 &  13.806001 &  2.036239 & -1.635468 &  0.018019 &  0.037600 \\
EOS3 &   8.695491 &  10.431351 &   9.821776 &  3.975509 & -2.615425 &  0.006394 &  0.039323 \\
EOS4 &   9.608190 &  11.957725 &  12.191950 &  3.117923 & -4.098400 &  0.000255 &  0.058744 \\
\hline
\end{tabular}
\caption{Parameters of the employed nuclear matter EOS: EOS1, EOS2, EOS3 and EOS4. The B and C are $b \times 10^3$, and $c \times 10^3$ respectively \cite{Malik:2023mnx}}. \label{tableI}

\end{table*}

\subsection{Dark matter EOS}\label{SubsecB2}
Similar to the Lagrangian of the nuclear model, one can apply the knowledge from the nuclear mean field approach to describe the Lagrangian for the fermionic dark matter sector. 
We consider the simplest dark matter Lagrangian with a single fermionic component ($\chi_D$) and we assume that a dark vector meson $V_D^{\mu}$ that couples to the conserved DM current through $g_{v d}\bar{\chi}_D\gamma_{\mu}\chi_DV^{\mu}_D$. 
 The dark matter model Lagrangian and the corresponding EOS in the mean field approximation is expressed as
\begin{eqnarray}
   {\cal{L}}_{\chi} &=&  \bar{\chi}_D \left[\gamma_{\mu}(i\partial^{\mu} - g_{v d}V^{\mu}_D) - m_{\chi}\right]\chi_D \nonumber\\ 
               &-& \frac{1}{4}V_{\mu\nu,D}V_D^{\mu\nu} + \frac{1}{2}m_{v d}^2V_{\mu,D}V_D^{\mu}
\end{eqnarray}

\begin{eqnarray}
\varepsilon_{\chi} = \frac{1}{\pi^2}\int_{0}^{k_D} dk~k^2\sqrt{k^2 + m_\chi^2} + \frac{1}{2} c_{\omega}^2\rho^2_D 
\end{eqnarray}

\begin{eqnarray}
P_{\chi} = \frac{1}{3\pi^2}\int_{0}^{k_D} dk \frac{k^4}{\sqrt{k^2 + m_{\chi}^2}} +\frac{1}{2} c_{\omega}^2\rho^2_D   
\end{eqnarray}

Here $c_{\omega }\equiv \frac{g_{vd }}{m_{vd }}$ and $m_{\chi}$ is the bare mass of fermionic dark matter. These two parameters along with the dark matter Fermi momenta determine the dark matter EOS. The dark matter Fermi momenta determines the accumulated dark matter density/mass fraction inside neutron stars. The properties of dark matter admixed neutron stars depend on the dark matter EOS along with dark matter mass fraction. Procedure to determine the dark matter EOS has been discussed in subsequent sections.

\begin{figure}
 \includegraphics[width=1.0\linewidth]{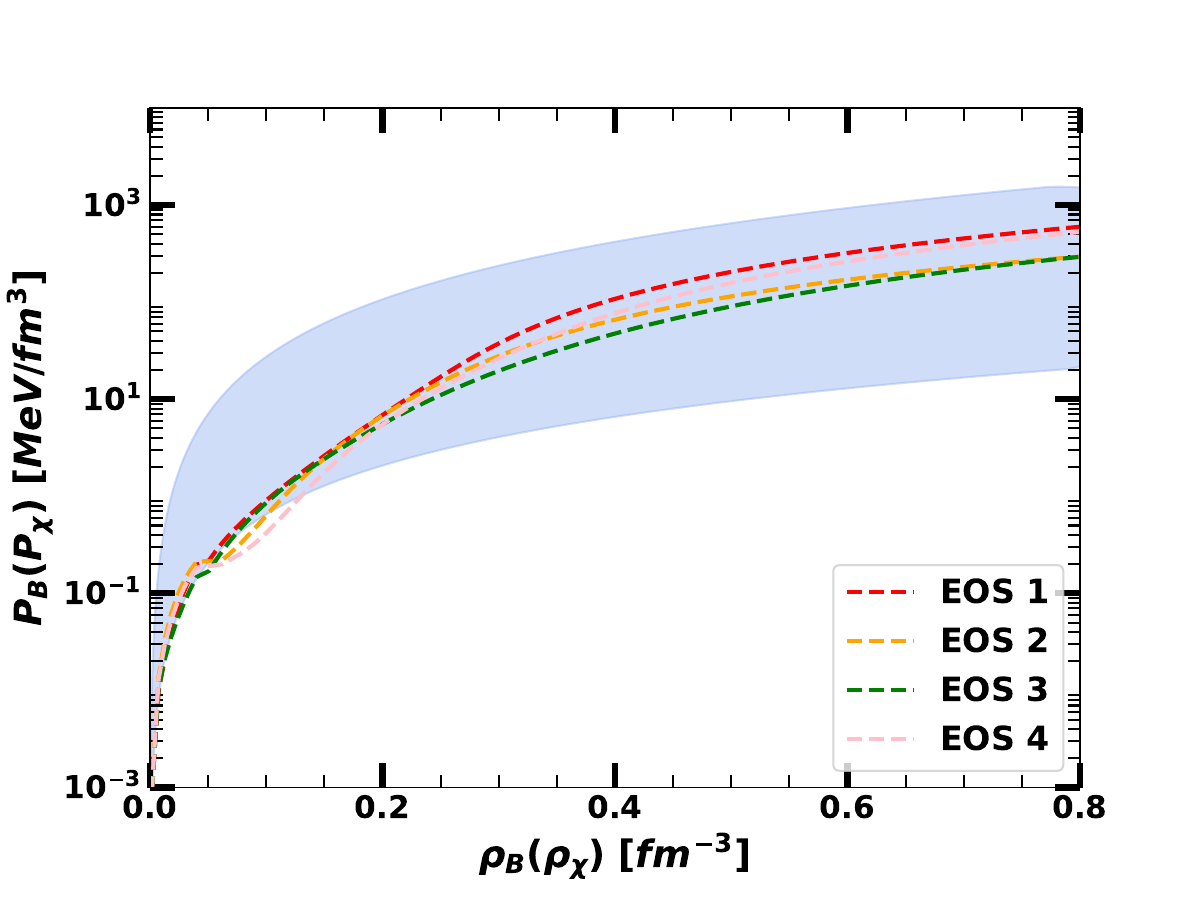}
 \caption{ The shaded blue domain represents the sampled dark matter EOS, i.e., the pressure ($P_{\chi}$) as a function of density ($\rho_{\chi}$). The colored dashed lines depict the nuclear matter EOS, i.e., the variation of baryonic matter pressure ($P_B$) with baryon number density ($\rho_B$).}
 \label{fig:eos_dark_matter}
 \end{figure}

\subsection{Two fluid Formalism}\label{SubsecB3}
We have employed a two-fluid Tolman-Oppenheimer-Volkoff (TOV) formalism to analyze the structure of neutron stars with a mixture of dark matter, referred to as Dark Matter Admixed Neutron Stars (DANSs)~\cite{Das:2020ecp}. Dark matter and baryonic matter are treated separately within this framework and interact solely through gravitational interaction. Consequently, each fluid follows its conservation of energy-momentum tensor.

To describe the combined effects of the two fluids, we introduce the total pressure $P(r)$ and total energy $\varepsilon(r)$, which can be expressed as the sum of the respective contributions from baryonic matter and dark matter:

\begin{eqnarray}
P(r) = P_{B}(r) + P_{\chi}(r)  \\ 
\varepsilon(r) = \varepsilon_{B}(r) + \varepsilon_{\chi}(r)
\end{eqnarray}

Here, the subscripts ``$B$" and ``$\chi$" represent the baryonic and dark matter components, respectively. The TOV equations governing the behavior of this two-fluid system are given by \cite{Das:2020ecp,Rutherford:2022xeb}: 

\begin{eqnarray}
\frac{dP_{B}}{dr} = -(P_B + \varepsilon_B)\frac{4\pi r^3(P_B+P_{\chi})+M(r)}{r(r-2M(r))}
\end{eqnarray}

\begin{eqnarray}
\frac{dP_{\chi}}{dr} = -(P_{\chi} + \varepsilon_{\chi})\frac{4\pi r^3(P_B+P_{\chi})+M(r)}{r(r-2M(r))}
\end{eqnarray}

\begin{eqnarray}
\frac{dM(r)}{dr} = 4\pi (\varepsilon_B + \varepsilon_\chi) r^2
\end{eqnarray}

When investigating the influence of admixed dark matter, it proves useful to define a dark matter mass fraction $F_{\chi}$~\cite{Rutherford:2022xeb}:
\begin{eqnarray}
F_{\chi} = \frac{M_{\chi}(R_\chi)}{M{(R)}}
\end{eqnarray}
Here, $M_{\chi}(R_{\chi})=4\pi\int_0^{R_\chi} r^2\varepsilon_{\chi} (r) dr$ represents the total accumulated dark matter gravitational mass within $R_{\chi}$, where the dark matter pressure reaches zero. Based on this DM mass fraction, it is possible to determine how much gravitational mass the DANS contributes to the star's total mass.

Besides mass and radius, neutron stars' tidal deformability plays a crucial role in their structural characteristics. The tidal gravitational field generated by their companion causes the two neutron stars in a binary neutron star system to undergo quadrupole deformations during the final stages of inspiration. As a result of the tidal forces exerted by the partner star of a neutron star, the magnitude of the deformation that occurs is described as tidal deformability, which quantifies the extent to which it distorts under those forces.

The dimensionless  tidal deformability is defined as 
\begin{eqnarray}
    \Lambda = 2/3 ~ k_2 ~ C^{-5}
\end{eqnarray}
where $C$ $(\equiv M/R)$ and $k_2$ are known as the compactness and Love number of the deformed star. $k_2$ for the two-fluid system can be obtained by solving the differential equation for radial perturbation, 
\begin{eqnarray}
 r \frac{d y(r)}{dr} + {y(r)}^2 + y(r) F(r) + r^2 Q(r) = 0,
  \end{eqnarray}
\begin{widetext}
\begin{eqnarray}
    F(r) = \frac{r-4 \pi r^3 \left( (\varepsilon_B(r)+\varepsilon_\chi(r)) - (P_B(r)+P_\chi(r)\right) }{r-2
M(r)}
\end{eqnarray}
\begin{eqnarray}
\nonumber Q(r) &=& \frac{4 \pi r \left(5 (\varepsilon_B(r)+\varepsilon_\chi(r)) +9 (P_B(r)+P_\chi(r)) +
\frac{\varepsilon_B(r) + P_B(r)}{\partial P_B(r)/\partial
\varepsilon_B(r)}+\frac{\varepsilon_\chi(r) + P_\chi(r)}{\partial P_\chi(r)/\partial
\varepsilon_\chi(r)} - \frac{6}{4 \pi r^2}\right)}{r-2M(r)} \\
&-&  4\left[\frac{M(r) + 4 \pi r^3
(P_B(r)+P_\chi(r))}{r^2\left(1-2M(r)/r\right)}\right]^2, \nonumber
\end{eqnarray}  
\end{widetext}
together with the two-fluid TOV equation with proper boundary conditions \cite{Das:2020ecp,Karkevandi:2021ygv}. 

\subsection{Sampling}
\label{SubsecB4}
To compute the structure of neutron stars (NS), we have utilized four distinct nucleonic equation-of-states (EOSs) and a diverse range of dark matter EOSs constructed through the Relativistic Mean Field (RMF) formalism, as detailed in sections \ref{SubsecB1} and \ref{SubsecB2} respectively. We have sampled a total of {50,000} dark matter parameter combinations, namely $c_{\omega}$, $m_{\chi}$, and $F_{\chi}$ from uniform distributions within the ranges specified in Table \ref{tab:dmt}. In Fig.\ref{fig:eos_dark_matter} we represent EOS for normal matter and dark matter components. There were 50K dark matter EOSs solved individually for each nucleonic EOS, resulting in a total of {200,000} mass-radius (M-R) calculations. To determine the dark matter EOS here we consider the dark matter with a mass range $0.5 ~\text{GeV}\leq m_{\chi}\leq 4.5~\text{GeV}$~\cite{Calmet:2020pub}, dark matter self-interaction measure in the range, $0.1 ~\text{fm}\leq c_{\omega} \leq 5 ~\text{fm}$~\cite{Xiang:2013xwa}, and dark matter mass fraction in the range $0\leq F_{\chi}\leq 25 \%$~\cite{Ciancarella:2020msu}. Note that the dark matter mass fraction crucially depends on the dark matter capture rate in
neutron stars. Depending upon the generic modeling of the nucleon-dark matter interaction inside the high-density region of neutron stars, the dark matter capture rate can be as high as $10^{43}$ GeV s$^{-1}$ for $1.0-10$ GeV dark matter particles. Due to the large capture rate, a typical one billion years old neutron star can accumulate a large fraction of dark matter~\cite{Bell:2020obw}.

\begin{table*}[]
\caption{Nuclear Saturation Properties - (i) For Symmetric Nuclear Matter - energy per nucleon $\varepsilon_0$, incompressibility coefficient $K_0$, and skewness $Q_0$; and (ii) For Symmetry Energy - symmetry energy at saturation $J_{\rm sym,0}$, its slope $L_{\rm sym,0}$; and Neutron Star Properties - maximum mass M$_{\rm max}$, radius $R_{\rm max}$, radius $R_{1.4}$ for 1.4 $M_\odot$ and $R_{2.08}$ for 2.08 $M_\odot$ neutron stars, tidal deformability $\Lambda_{1.4}$ for 1.4 solar mass neutron stars, the square of speed-of-sound $c_s^2$ at the center of maximum mass neutron stars, the neutron star mass at which the direct Urca process occurs $M_{\rm dUrca}$, and the direct Urca density $\rho_{B,x}$ for $x \in [1.4, 1.6, 1.8]$ solar mass neutron stars. \label{tableII}}
\setlength{\tabcolsep}{3.0pt}
\renewcommand{\arraystretch}{1.2}
\begin{tabular}{cccccccccccccccccccccc}
\hline \hline 
\multirow{3}{*}{EOS} & \multicolumn{8}{c}{NMP}                                                                                               & \multicolumn{9}{c}{NS}                                                                                                                                                                                                                                         \\ \cline{2-18} 
                     & $\rho_0$      & $\varepsilon_0$ & $K_0$ & $Q_0$ & $J_{\rm sym,0}$ & $L_{\rm sym,0}$ & $M_{\rm max}$   & $R_{\rm max}$ & $R_{1.4}$ & $R_{2.08}$ & $\Lambda_{1.4}$ & $c_s^2$     & \multicolumn{1}{c}{$M_{\rm dUrca}$} & \multicolumn{1}{c}{$\rho_{\rm dUrca}$} & \multicolumn{1}{c}{$\rho_{\rm B,1.4}$} & \multicolumn{1}{c}{$\rho_{\rm B,1.6}$} & \multicolumn{1}{c}{$\rho_{\rm B,1.8}$} \\
                     & {[}fm$^{-3}${]} & \multicolumn{5}{c}{{[}MeV{]}}                                       & {[}M$_\odot${]} & \multicolumn{3}{c}{{[}km{]}}           & {[}$...${]}   & {[}$c^2${]} & \multicolumn{1}{c}{{[}M$_\odot${]}} & \multicolumn{4}{c}{{[}fm$^{-3}${]}}                                                                                                                                 \\ \hline
EOS1                & 0.155         & -16.08          & 177   & -74   & 33              & 64              & 2.74            & 13.03         & 13.78     & 14.04      & 844           & 0.713       & 2.06                                & 0.366                                  & 0.298                                  & 0.316                                  & 0.336                                  \\
EOS2                & 0.154         & -15.72          & 190   & 614   & 32              & 60              & 2.20            & 12.16         & 13.36     & 13.00      & 709           & 0.414       & 1.83                                & 0.443                                  & 0.344                                  & 0.382                                  & 0.432                                  \\
EOS3                & 0.157         & -16.24          & 260   & -400  & 32              & 57              & 2.10            & 11.08         & 12.55     & 11.53      & 462           & 0.543       & 2.07                                & 0.829                                  & 0.432                                  & 0.491                                  & 0.570                                  \\
EOS4                & 0.156         & -16.12          & 216   & -339  & 29              & 42              & 2.56            & 12.13         & 12.95     & 13.14      & 638           & 0.767       & 2.55                                & 0.747                                  & 0.345                                  & 0.370                                  & 0.399                                  \\ \hline
\end{tabular}
\end{table*}

\begin{table}[htp]
\caption{The prior set  for dark matter model parameters.}
\setlength{\tabcolsep}{11.0pt}
      \renewcommand{\arraystretch}{1.4}
\begin{tabular}{cccccc}
\hline \hline 
\multicolumn{2}{c}{$m_{\chi}$} & \multicolumn{2}{c}{$c_{\omega}$} & \multicolumn{2}{c}{$F_{\chi}$} \\
\multicolumn{2}{c}{GeV}        & \multicolumn{2}{c}{fm}   & \multicolumn{2}{c}{\%}        \\ \hline
min           & max            & min             & max            & min           & max          \\ \hline
0.5           & 4.5           & 0.1               & 5             & 0             & 25         \\ \hline
\end{tabular}
\label{tab:dmt}
\end{table}

\section{Results and Discussion} \label{results}
We aim to investigate whether NS observational properties such as mass, radius, and tidal deformability, can uniquely constrain the dark matter model parameters. To explore this, we have employed a two-fluid scenario to calculate neutron star properties \ref{SubsecB3}. For the nuclear matter, we utilize four realistic EOS, namely EOS1, EOS2, EOS3, and EOS4 (see Sec. \ref{SubsecB1} for details). In the dark matter sector, a total of 50K dark matter EOSs were sampled within the reasonable prior range (see Table \ref{tab:dmt}, Sec. \ref{SubsecB2} and Sec. \ref{SubsecB4} for details). These large combinations of nuclear matter and dark matter EOS give rise to 200K mass-radius curves by solving the two-fluid TOV equations. All of these mass-radius relations are not practically relevant as we use the filter that NS must have a mass greater than {1.9 M$_\odot$, set by the pulsars  PSR J0348+0432  and  PSR J0740+6620  within $\sim 3 \sigma$}. We also consider that the dark matter admixed neutron stars only produce non-halo configurations, which means the dark matter admixed radius is smaller than the luminous radius. Following the application of this filter, the remaining values of samples for EOS1, EOS2, EOS3, and EOS4 are 25K, 14K, 10K, and 20K, respectively. {The median values, along with the lower and upper bounds of the 90\% confidence intervals (CI), for various neutron star properties for these sets are listed in Table \ref{propall}. These properties include the maximum mass of neutron stars (M$_{\rm max}$), the total radius (R$_{t, x}$) for values of $x$ in the range of [1.2, 1.4, 1.6, 1.8, 2.0], the dimensionless tidal deformability ($\Lambda_{x}$) for $x$ in the range of [1.4, 1.6, 1.8], the fraction of dark matter energy density over nuclear matter ($f_{d,x}$), and the nuclear matter baryon density ($\rho_{B,x}$) for $x$ in the range of [1.4, 1.6, 2.0].}

\begin{figure*}[htp]
    \centering
    \includegraphics[width=0.95\textwidth]{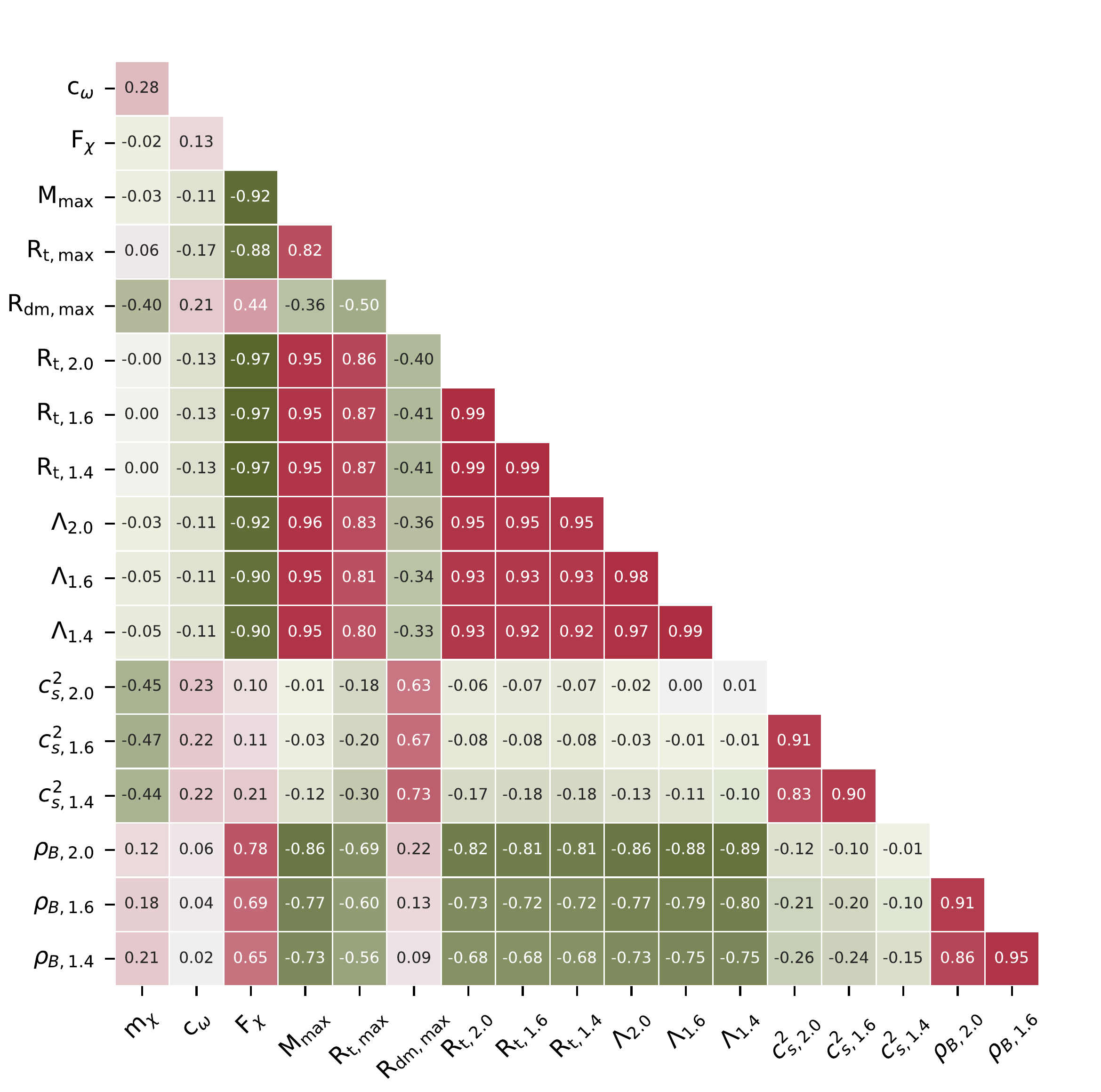}
    \caption{With only one nuclear EOS, namely EOS1, we compute the Kendall rank correlation coefficients linking various dark matter parameters to neutron star properties, and with the entire set of dark matter EOS sets after applying the filter (see text).}
     \label{fig:cor_eos1}
\end{figure*}

\begin{figure*}[htp]
    \centering
    \includegraphics[width=0.95 \textwidth]{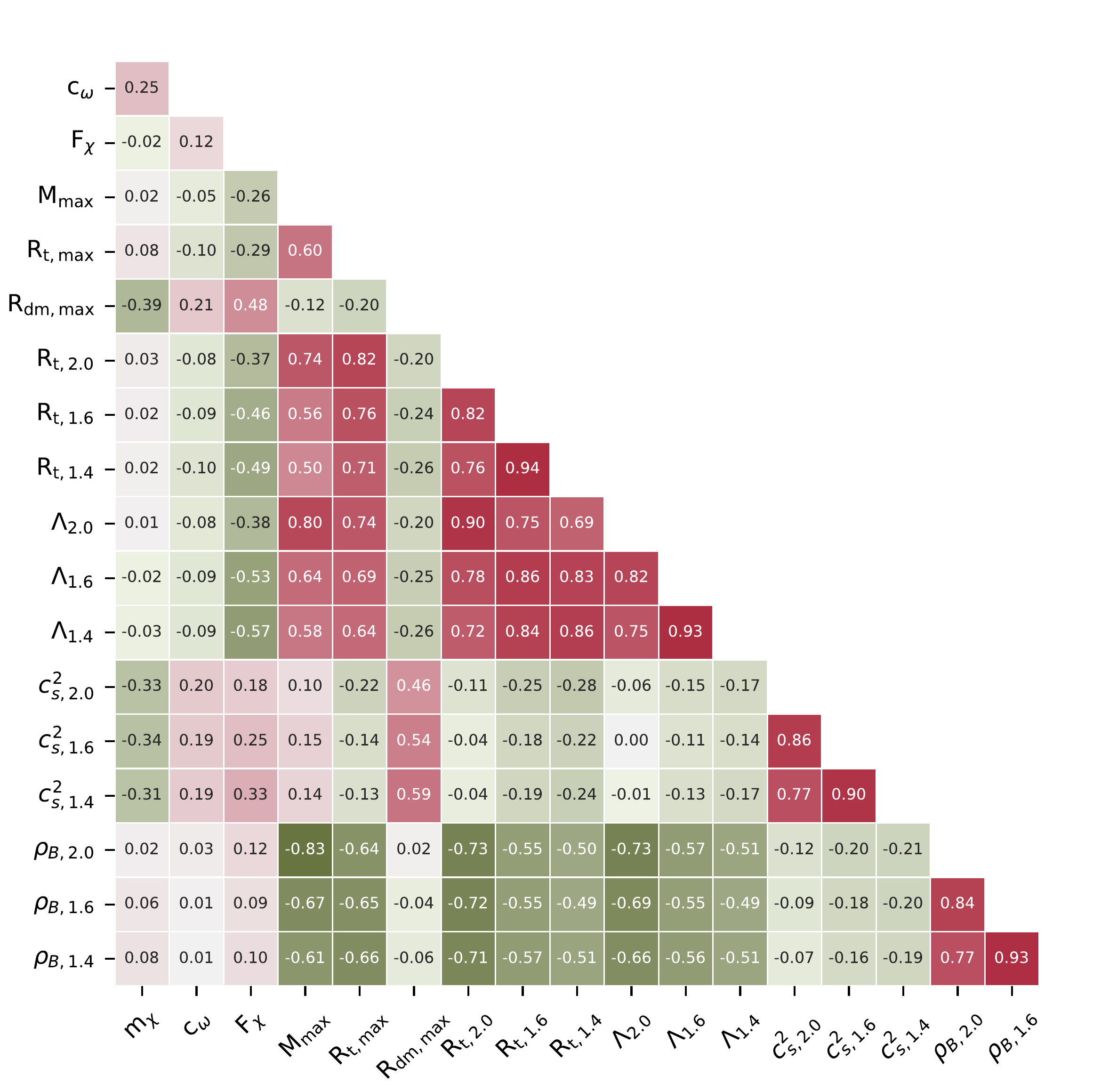}
    \caption{Same as figure \ref{fig:cor_eos1} but with the data where all four nuclear EOS are employed.}
     \label{fig:cor_mix}
\end{figure*}
To study the relation between dark matter parameters and various neutron star properties, we employ the Kendall rank correlation coefficient analysis. The Kendall rank correlation can be considered a non-parametric test that allows us to quantify the strength of dependence between two variables. Unlike Pearson correlation, it accounts for non-linearity in the correlation, making it suitable for analyzing relationships that may not follow a linear pattern. Our strategy involves first selecting a fixed nuclear EOS, and then assessing the correlation coefficients using the chosen set of dark matter EOS. In this way, we can explore the uncertainty only in the dark matter sector. 
Lastly, we explore the combined effects of all four combinations of nuclear equations and the entire set of dark matter equation-of-states. This approach helps us to comprehend the impact of uncertainty in the nuclear EOS on these correlations.
In Fig.\ref{fig:cor_eos1} we plot the Kendall rank correlation coefficients, where 
we consider the nuclear EOS1 and the entire set of filtered dark matter equation-of-state sets, i.e., with NS having a maximum mass above 1.9 $M_{\odot}$ and NS having to be non-halo (as mentioned earlier). As portrayed in Fig. \ref{fig:cor_eos1}, Fig. \ref{fig:cor_mix} includes the complete data set with all four nuclear EOSs employed. Please note that in this context, the subscript $`t'$ denotes the total radius of the neutron star, while the subscript $`dm'$ represents the radius specifically associated with the {admixed} dark matter.

The following comments are in order:
\begin{itemize}
   \item We have observed in Fig.\ref{fig:cor_eos1} a strong negative correlation $\sim$ 0.9 between the dark matter mass fraction, $F_\chi$, and the maximum gravitational mass of a neutron star (NS). Additionally, there is a notable correlation $\sim$ 0.9  between $F_\chi$ and the radius as well as the tidal deformability of NS at masses of 1.4, 1.6, and 2.0 $M_{\odot}$ respectively. However in Fig.\ref{fig:cor_mix}, when incorporating uncertainty in the nuclear sector and considering all four nuclear matter equation-of-states (EOSs) alongside the sampled dark matter EOSs, the previously mentioned correlation disappears. 
   
   \item Furthermore, our findings have revealed a notable positive correlation between $F_\chi$ and the central baryonic density $\rho_B$ (at different NS masses 1.4, 1.6, and 2.0 $M_{\odot}$) for EOS1. However, once again when we account for the uncertainties associated with nuclear matter, as can be seen in Fig.\ref{fig:cor_mix}, those correlations disappear.
   
   \item {Nonetheless, the correlations between radii, tidal deformability, mass, and central baryon density persist even when considering uncertainties in the nuclear sector, i.e., including the entire dark matter admixed set for all nuclear EOS. It is worth noting that there is a robust and strong correlation between the central baryon density and the star radius even when dark matter is not considered \cite{Malik:2023mnx, Jiang:2022tps}. However, this correlation is found to break in the case of modified gravity \cite{Nobleson:2023wca}. This interestingly allows us to distinguish between the effects of modified gravity and dark matter in neutron stars (NS).}
\end{itemize}

\begin{figure*}[htp]
    \centering
    \includegraphics[width=0.45\textwidth]{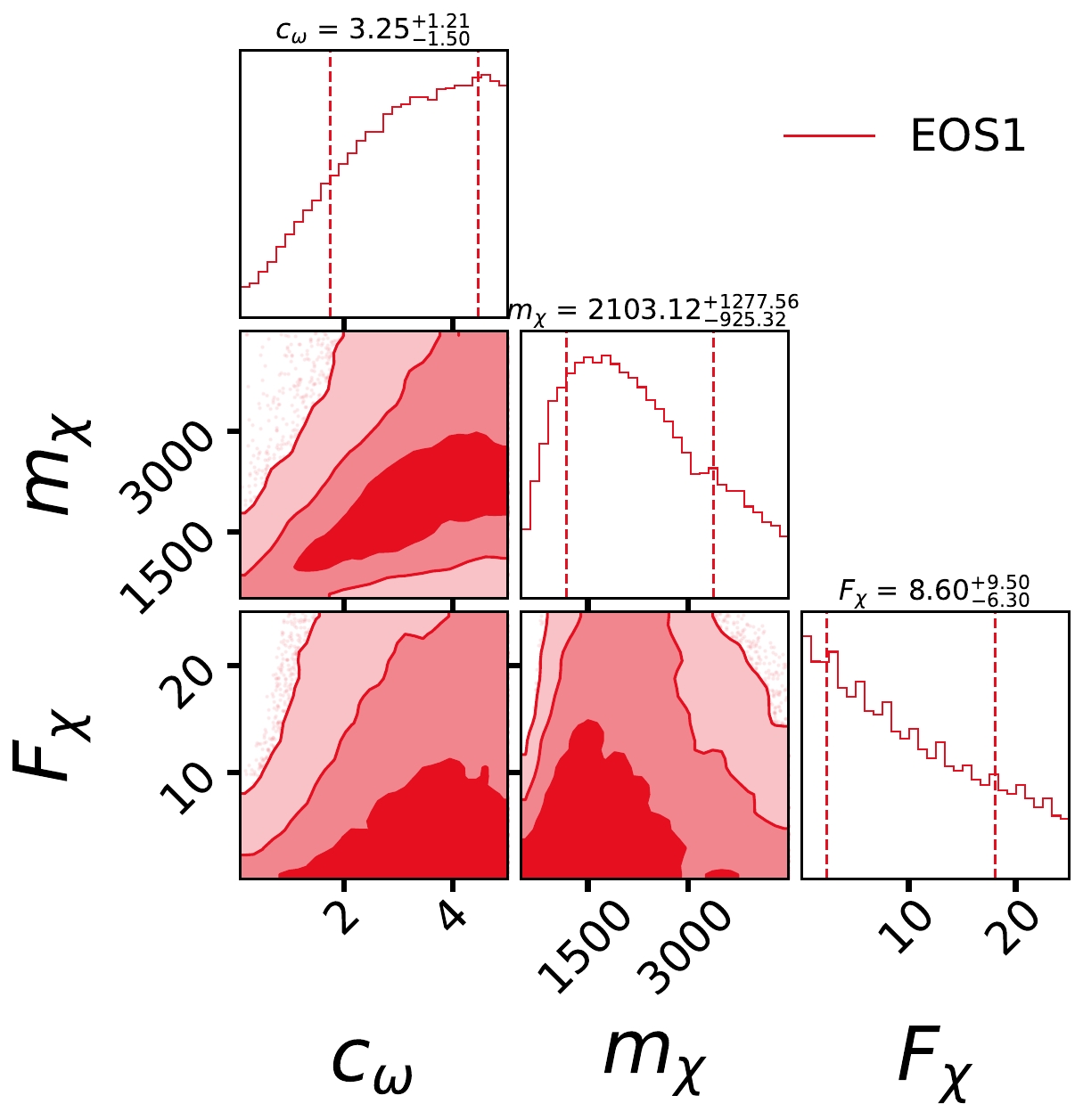}
    \includegraphics[width=0.45\textwidth]{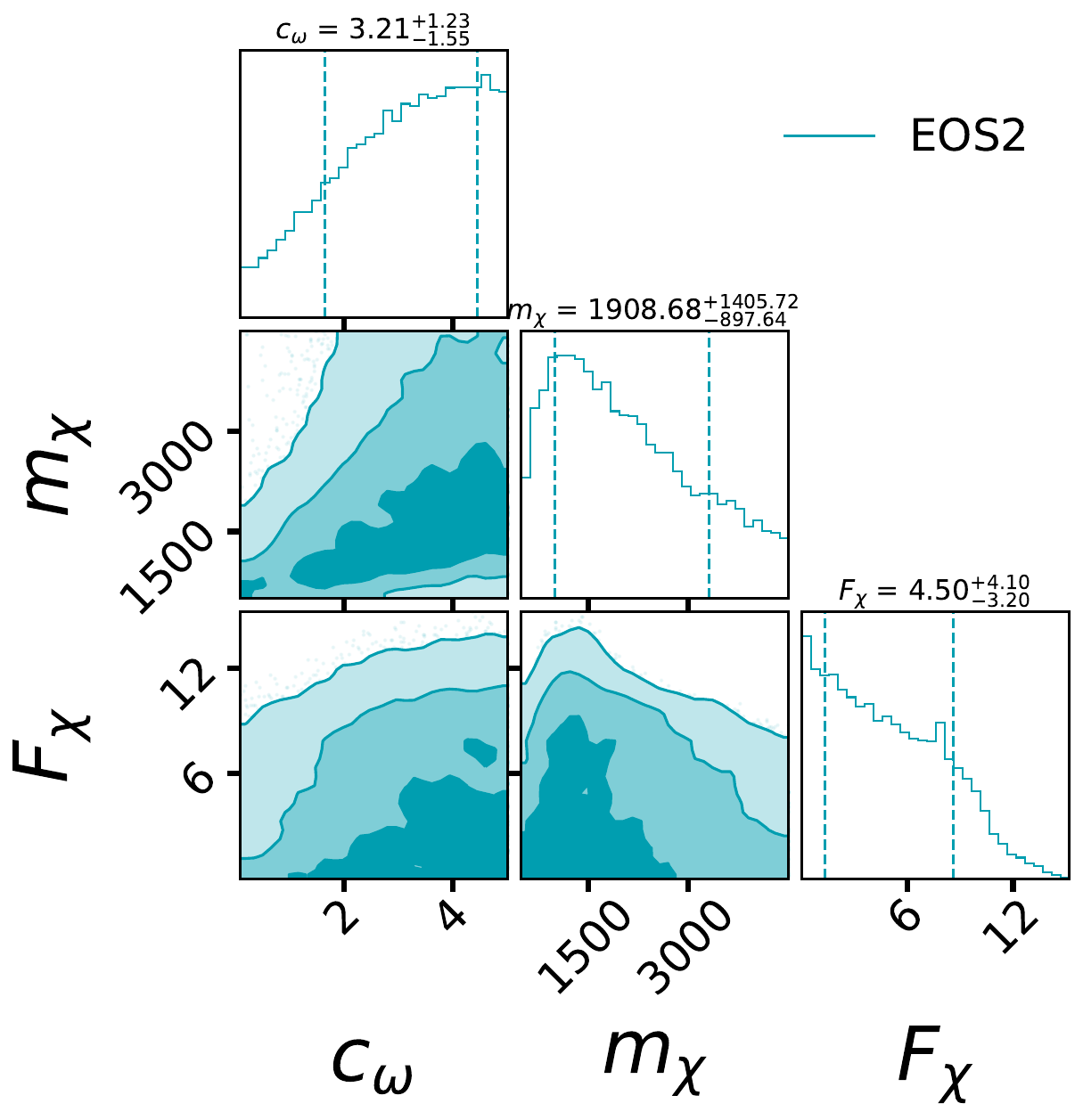}
    \includegraphics[width=0.45\textwidth]{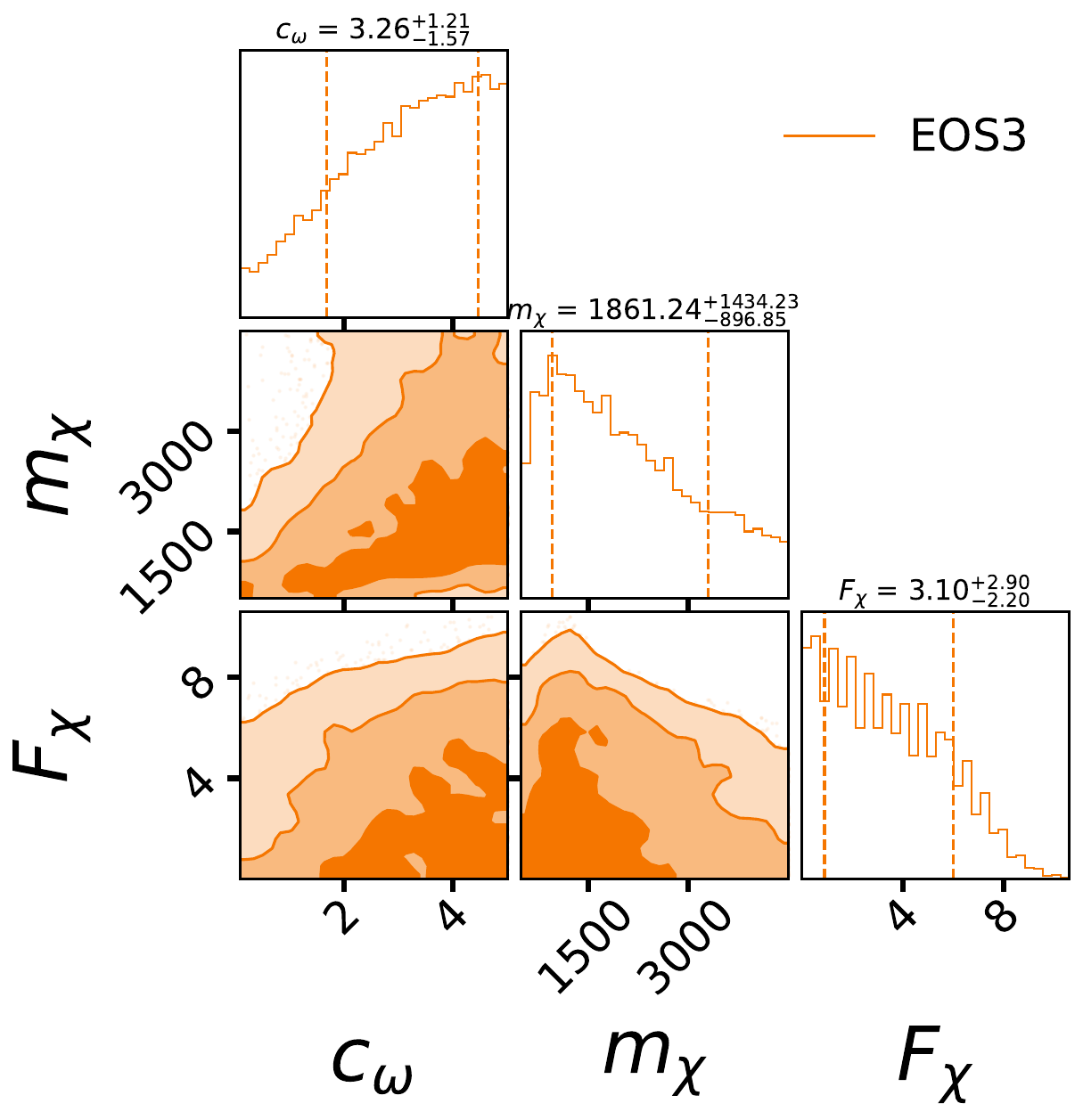}
    \includegraphics[width=0.45\textwidth]{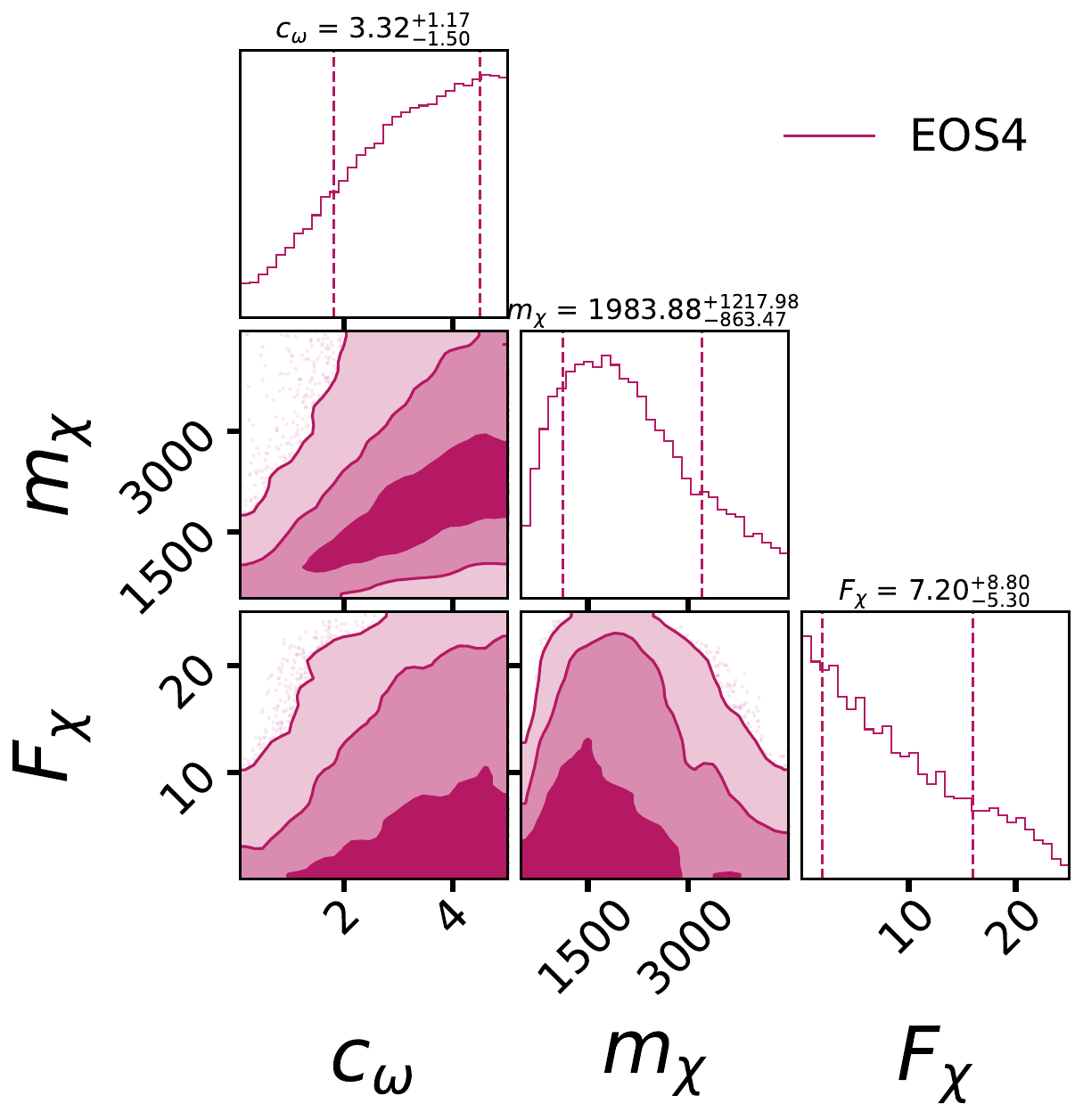}
    \caption{The distribution of dark matter parameters, i.e., $c_{\omega}$, $m_{\chi}$ (in MeV units), and $F_{\chi}$ for the prior set mentioned in Table \ref{tab:dmt} after applying the filter that NS must have a mass greater than 1.9 M$_\odot$. }
     \label{fig:para}
\end{figure*}
Figure \ref{fig:para} displays the corner plot featuring the dark matter parameters $c_\omega$, $m_\chi$, and $F_\chi$ corresponding to EOS1, EOS2, EOS3, and EOS4 after applying the filtration process. A corner plot is a visualization used to explore the multivariate distribution of parameters in a dataset. It allows us to observe the relationships and correlations between different variables simultaneously, providing valuable insights into the underlying data structure. When considering a fixed nuclear EOS and only varying the dark matter EOS within the two-fluid formalism, there exists a direct correlation between the fraction of dark matter $F_\chi$ and $M_{\text{max}}$, as depicted in Fig. \ref{fig:cor_eos1}. Consequently, the maximum mass constraints of 1.9 $M_{\odot}$ depend on the softness or stiffness of the EOS for each individual nuclear matter EOS. Each nuclear matter EOS is capable of sustaining a different percentage of dark matter $F_\chi$. For example, the softest EOS3 can sustain only up to $\approx$ 10$\%$ of dark matter. whereas EOS1, EOS2 and EOS4   can sustain $\approx$ 24 $\%$, 14$\%$ and 22$\%$ of dark matter respectively. Therefore, depending on the stiffness of the employed EOS, we can observe variations in the percentage of dark matter fraction $ F_{\chi}$ranging from 0$\%$ to 25$\%$.

\begin{figure}[htb!]
    \centering
    \includegraphics[width=0.5\textwidth]{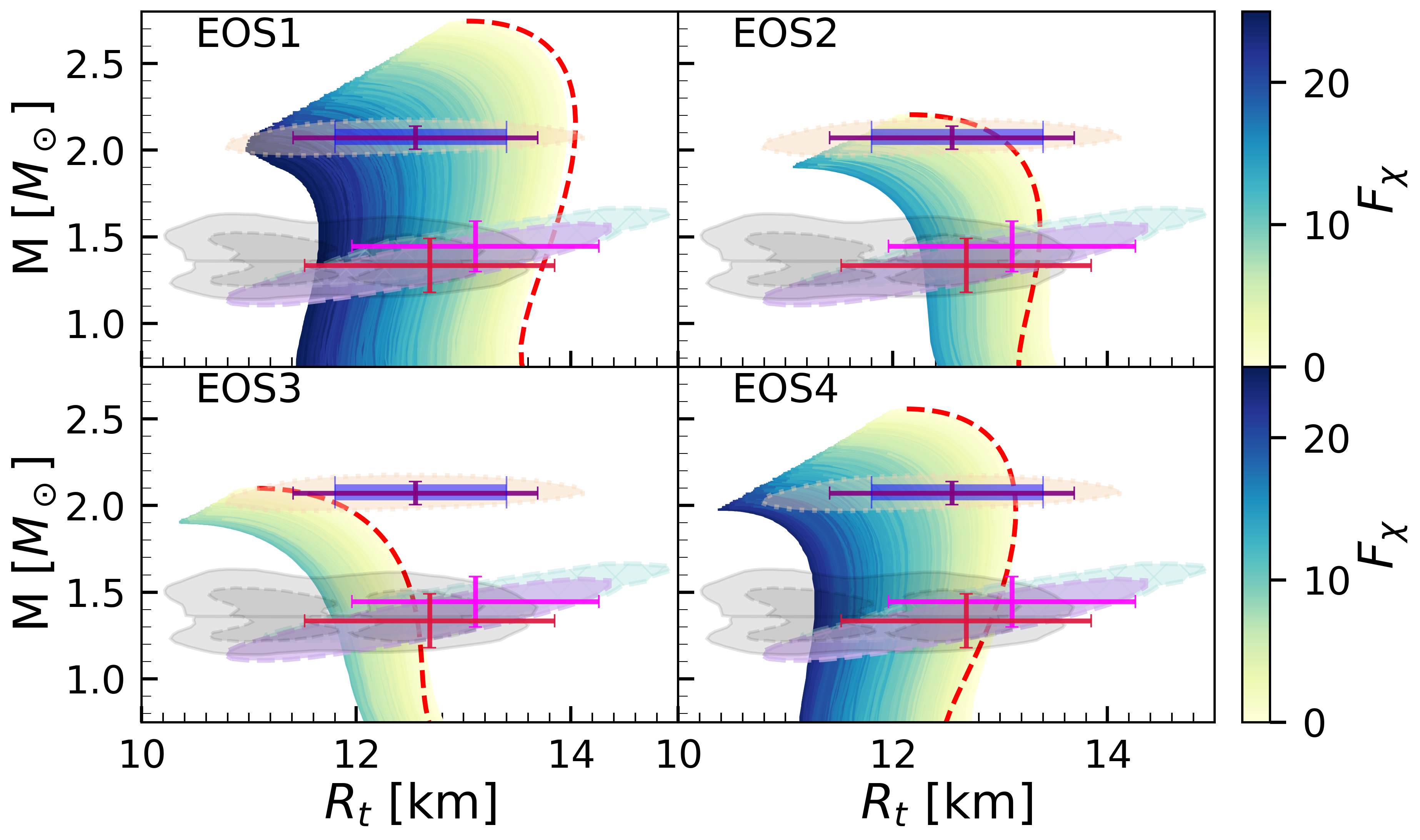}  
    \caption{The domain of neutron star (NS) mass-radius using a two-fluid scenario, considering our entire set of dark matter EOSs in conjunction with different nuclear EOSs. The {vertical} color bar with the panel depicts the dark matter mass fraction $F_{\chi}$ from 0 to 25 $\%$. The dashed lines on each plot correspond to NS properties computed using only nuclear EOS in single fluid TOV, without dark matter. We compare the M-R domains with current observational constraints. The gray region depicts the constraints from the binary components of GW170817 \cite{LIGOScientific:2018hze}, along with their 90$\%$ and 50$\%$ credible intervals(CI). The 1$\sigma$(68$\%$) CI  for the 2D posterior distribution in the mass-radii domain for millisecond pulsar PSRJ0030+0451 (cyan and yellow) \cite{Riley:2019yda, Miller:2019cac} as well as PSRJ0740+6620 (violate) \cite{Riley:2021pdl, Miller:2021qha} from the NICER x-ray data are also shown.}
    \label{fig:M_R_per}
\end{figure}
In Figure \ref{fig:M_R_per}, we demonstrate the outcomes of mass-radius calculations obtained through a collection of EOSs ensemble. The upper panels illustrate the results for EOS1 and EOS2, whereas the lower panels correspond to EOS3 and EOS4 respectively. The {vertical} color bar, which ranges from 0$\%$ to 25$\%$, visually illustrates the spread of dark matter mass fractions. It provides insight into the resulting variations in the mass-radius curve influenced by the parameter $F_{\chi}$. Interestingly, when the percentage of $F_{\chi}$ increases, it noticeably leads to a decrease in the maximum mass of neutron stars. The dashed lines present in each plot correspond to the properties of neutron stars computed exclusively based on the nuclear EOS, without taking into account the influence of dark matter. To assess the validity of our findings, we compare them with recent observational constraints, represented by skin lines. These constraints encompass the binary components of GW170817 \cite{LIGOScientific:2018hze}, along with their corresponding 90$\%$ and 50$\%$ credible intervals (CI). Furthermore, we illustrate the 1$\sigma$ (68$\%$) CI for the two-dimensional posterior distribution in the mass radii domain obtained from NICER x-ray data for the millisecond pulsars PSRJ0030+0451 (cyan and yellow) and PSRJ0740+6620 (violet). The horizontal (radius) and vertical (mass) error bars reflect the 1$\sigma$ credible interval derived from the 1-dimensional marginalized posterior distribution of the same NICER data. From this figure, one may observe that as the percentage of the dark matter component increases, both the mass and radius decrease for different neutron star mass sequences. It is worth noting that the current observational constraints on mass and radius, whether from NICER or GW observations, are not able to precisely determine the dark matter fraction $F_{\chi}$. Therefore, using the robust investigation presented in this Fig.\ref{fig:M_R_per}, we suggest that the dark matter fraction can be as high as 25$\%$, when 1.9 M$_\odot$ NS maximum mass constraint is imposed.

\begin{figure}[t!]
    \centering
    \includegraphics[width=0.48\textwidth]{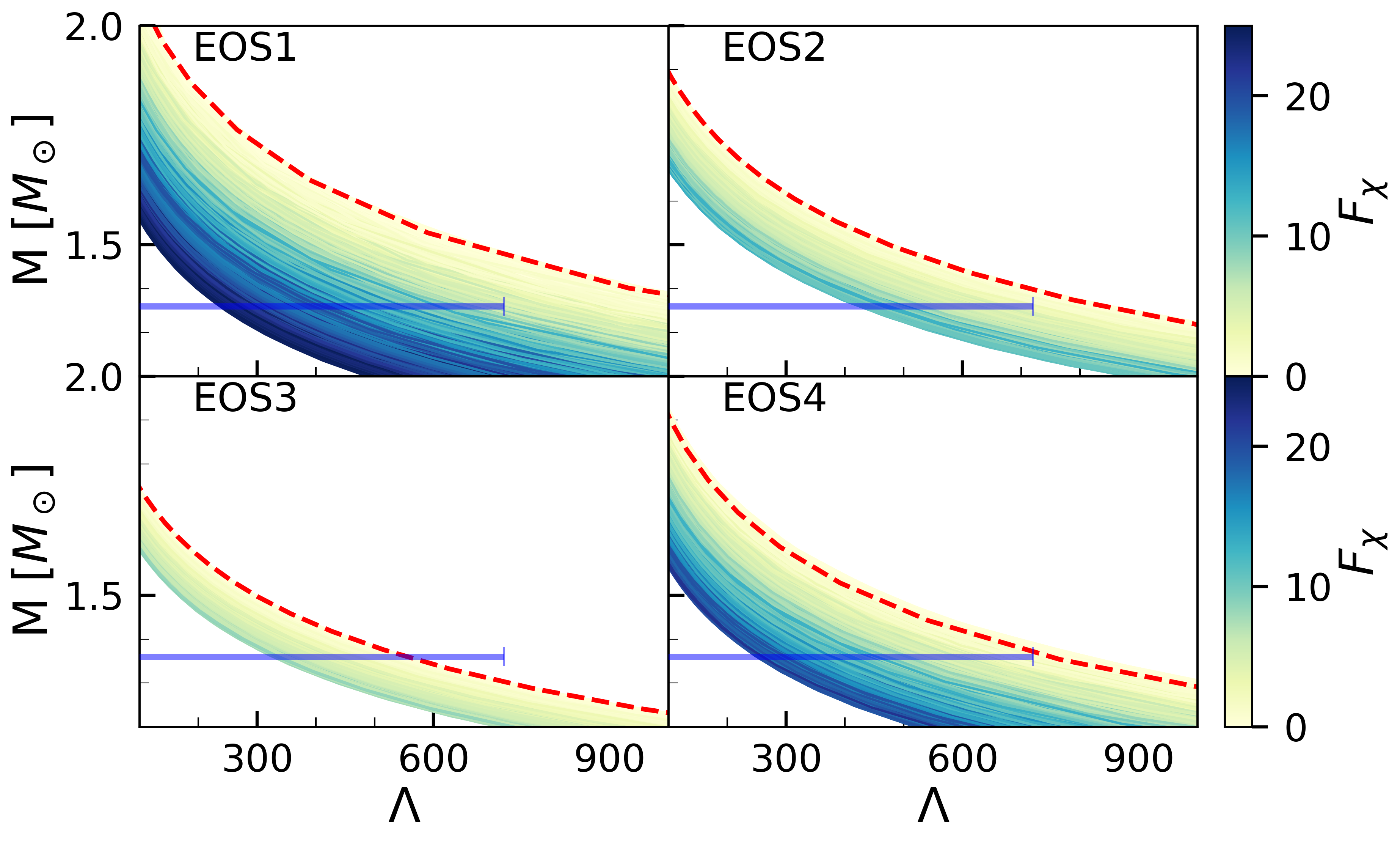}
    \caption{The graphical representation illustrates the relationship between tidal deformability ($\Lambda$) and the mass ($M_{\odot}$) of neutron stars (NS). Meanwhile, the dashed lines depicted in each plot correspond to the computed properties of neutron stars solely based on the nuclear EOS in single fluid TOV, incorporating the observational constraint (blue bars) depict the tidal deformability
    at 1.36 M$_\odot$ \cite{LIGOScientific:2018hze}. Additionally, the color gradient on the side denotes the proportional dark matter mass fraction $F_{\chi}$, encompassing values from 0$\%$ to 25$\%$.}
    \label{fig:L1L2}
\end{figure}
Figure \ref{fig:L1L2} illustrates the relation between the dimensionless tidal deformability ($\Lambda$) and the mass of neutron stars (NS) for different nuclear equation-of-states (EOSs) in separate panels. The dashed lines represent the properties of NS computed solely using the nuclear EOS in a single fluid TOV calculation, excluding the presence of dark matter. The color bar on the side indicates the dark matter mass fraction ($F_{\chi}$), with the color tone varying from yellow to blue, representing DM mass fractions ranging from 0$\%$ to 25$\%$. From the figure, it is clear that the inclusion of dark matter leads to a decrease in tidal deformability for all masses, the same as obtained in the previous figure for the radius. 
{As can be seen from the figure, dimensionless tidal deformability of different NS masses was negatively correlated with dark matter mass fraction $F_{\chi}$.} The inclusion of observational constraints from GW170817 is represented by the blue bars, depicting the tidal deformability at 1.36 M$_\odot$, $\Lambda_{1.36}<720$, \cite{LIGOScientific:2018hze}. 
The inclusion of dark matter could potentially lead to a reduction in the higher tidal deformability attributed to the stiff nuclear EOS. This similarity holds even for bosonic dark matter when employing a two-fluid approach, as demonstrated in previous studies Refs. \cite{Ivanytskyi:2019wxd, Rutherford:2022xeb}. The same behavior was obtained with a fermionic dark matter model based on RMF description incorporating short-range correlations  within a single fluid approach \cite{Lourenco:2022fmf}, and with  the linear sigma-omega fermionic dark matter model together with a two-fluid approach  \cite{Das:2020ecp}.

\begin{figure}
    \centering
    \includegraphics[width=0.48\textwidth]{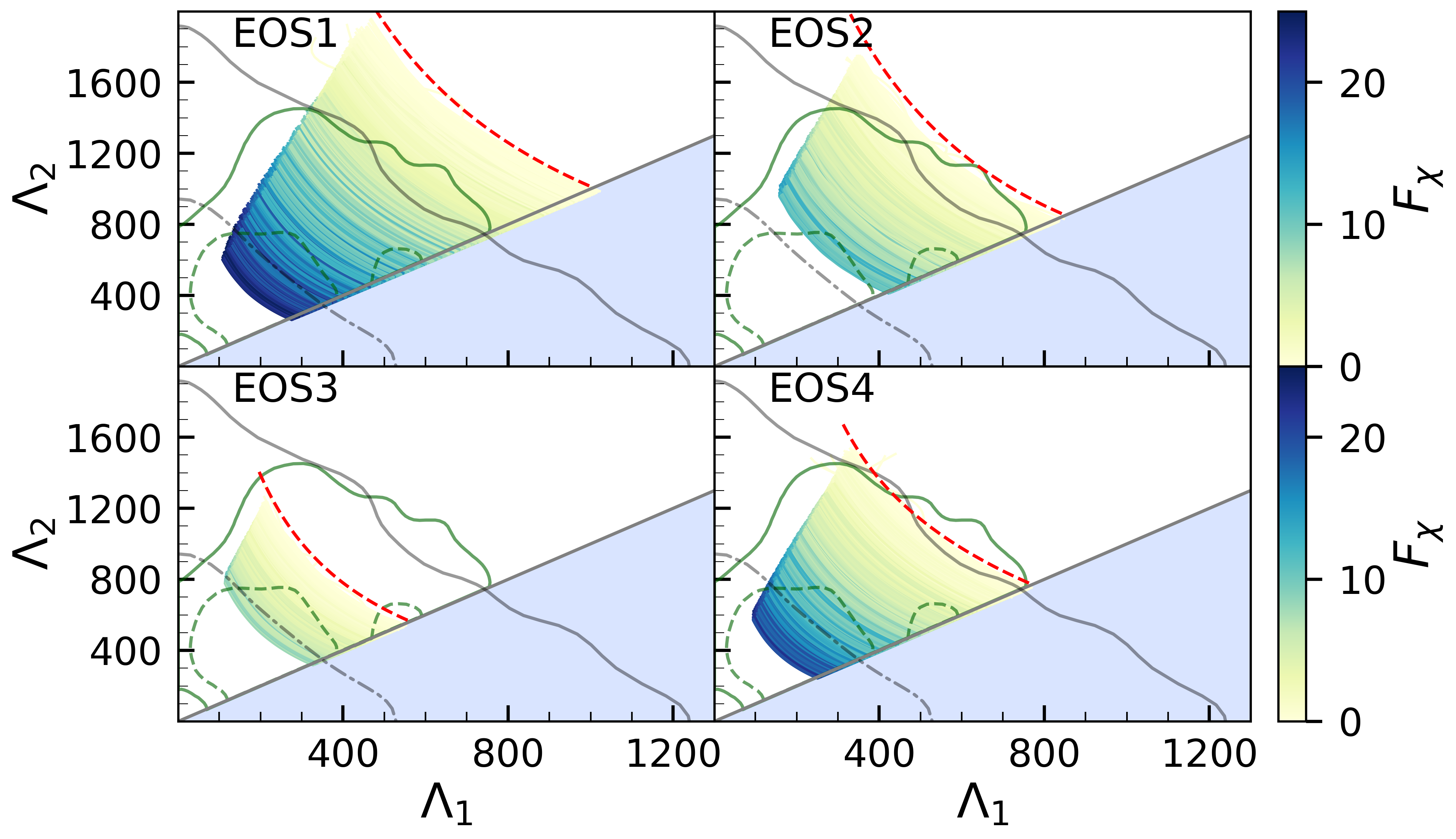}
    \caption{The graphical representation of $\Lambda_1$, and $\Lambda_2$ with a fraction of dark matter $F_{\chi}$ from 0 to 25$\%$, where $\Lambda_1$, and $\Lambda_2$ are the dimensionless tidal deformability parameters of the binary neutron star merger from the GW170817 event, using the observed chirp mass of $M_{chirp}$ = 1.186 $M_{\odot}$. The  green and gray  solid (dashed) lines represent the 
    90{$\%$} (50{$\%$}) CI from the marginalized posterior for the tidal deformabilities of the two binary components of GW170817 using  a parametrized EOS, with (green) and without (gray) 
     a maximum mass of 1.97 $M_\odot$ requirement.} 
    \label{fig:lam1_lam2}
\end{figure}
Fig.\ref{fig:lam1_lam2} illustrates the representation of $\Lambda_1$ and $\Lambda_2$, the dimensionless tidal deformability parameters obtained with nuclear matter EOS: EOS1, EOS2, EOS3, and EOS4 for the binary neutron star merger event GW170817. 
For this calculation, we have fixed the chirp mass ($M_{\rm chirp}$)  1.186 $M_{\odot}$ which is observed in the GW170817 event. In the plot, for the comparison we have included the constraints in the gray solid (dashed) line corresponding to the 90$\%$ (50$\%$) confidence interval (CI) obtained from the marginalized posterior, which represents the tidal deformability of the two binary components of neutron star merger event GW170817. Furthermore, the green solid (dashed) lines depict the 90$\%$ (50$\%$) CI derived from the marginalized posterior, indicating the tidal deformability of the two binary components of GW170817 based on an equation-of-state which is parameterized with a requirement that the maximum mass of at least 1.97 $M_{\odot}$. Here it can be seen that, for  EOS1 in the absence of any dark matter component lies outside the boundary of  observational constraints, but in the presence of dark matter as $F_{\chi}$ increases this comes inside the boundaries which results that the stiff nuclear EOS with admixed dark matter, comes inside the boundaries defined by the constraints on tidal deformability. As it is discussed in Fig. \ref{fig:para} EOS2 and EOS3 can only sustain upto $\approx$ 14$\%$ and 10$\%$ of dark matter respectively, if the 1.9 $M_\odot$ constraint is imposed. As a consequence, the acceptable $\Lambda_1$-$\Lambda_2$ domain is quite small, whereas for EOS1  and EOS4 the domain is wider because these EOS can sustain, respectively, 24$\%$ and 22$\%$ of dark matter.

\begin{figure}
    \centering
    \includegraphics[width=0.48\textwidth]{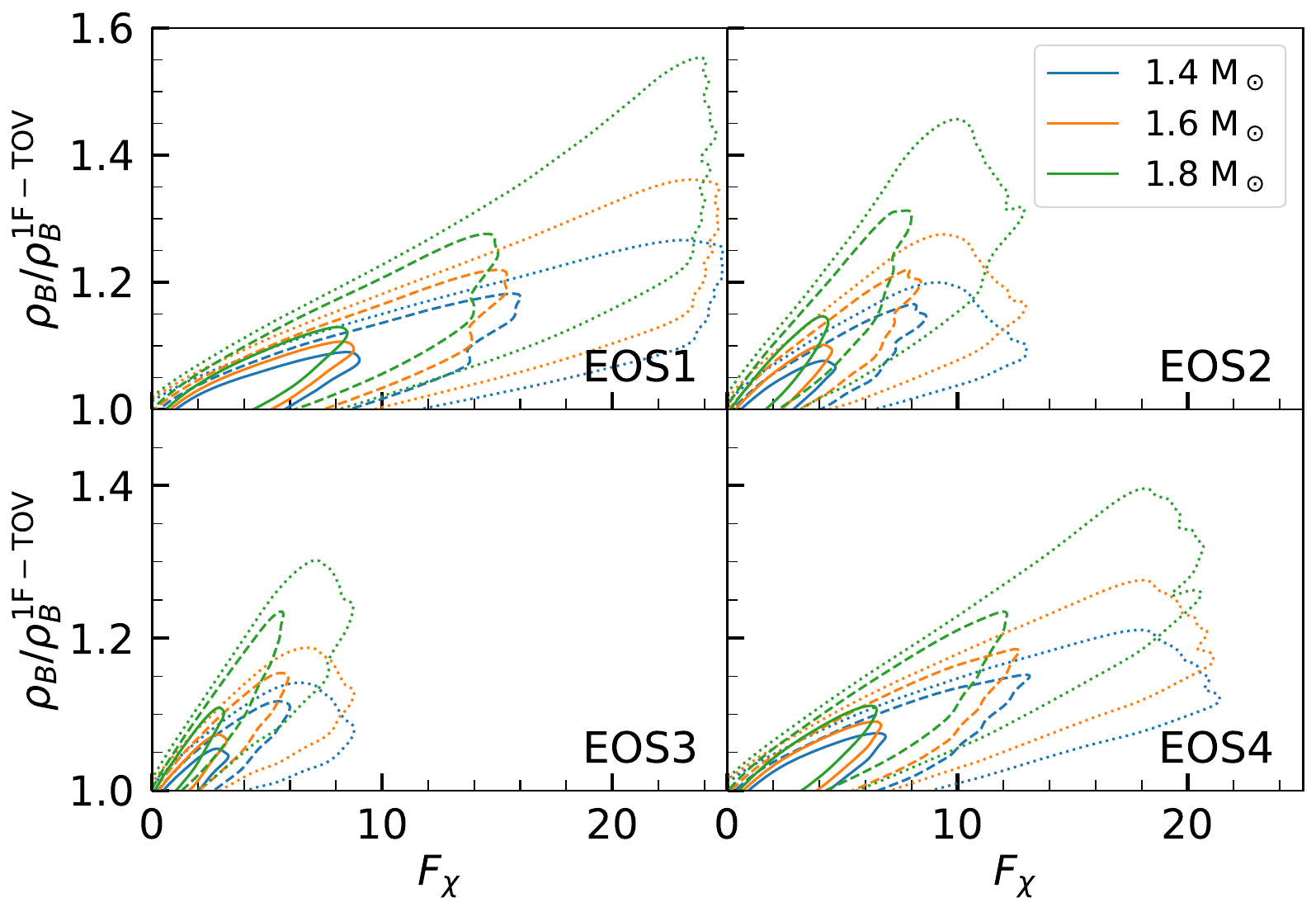}
    \caption{The central baryonic density $\rho_B$/$\rho_B^{1F-TOV}$ as a function of dark matter mass fraction $F_{\chi}$ for NS masses equal to 1.4 $M_{\odot}$(blue), 1.6  $M_{\odot}$(orange), 1.8  $M_{\odot}$(green). The top panels are for EOS1 and EOS2, whereas the bottom panels are for EOS3 and EOS4. The solid line in each panel represents 68$\%$ confidence interval whereas dashed and dotted lines represent 95$\%$ and 99$\%$, respectively.}
    \label{fig:rhob_fchi}
\end{figure}

\begin{figure}
    \centering
    \includegraphics[width=0.48\textwidth]{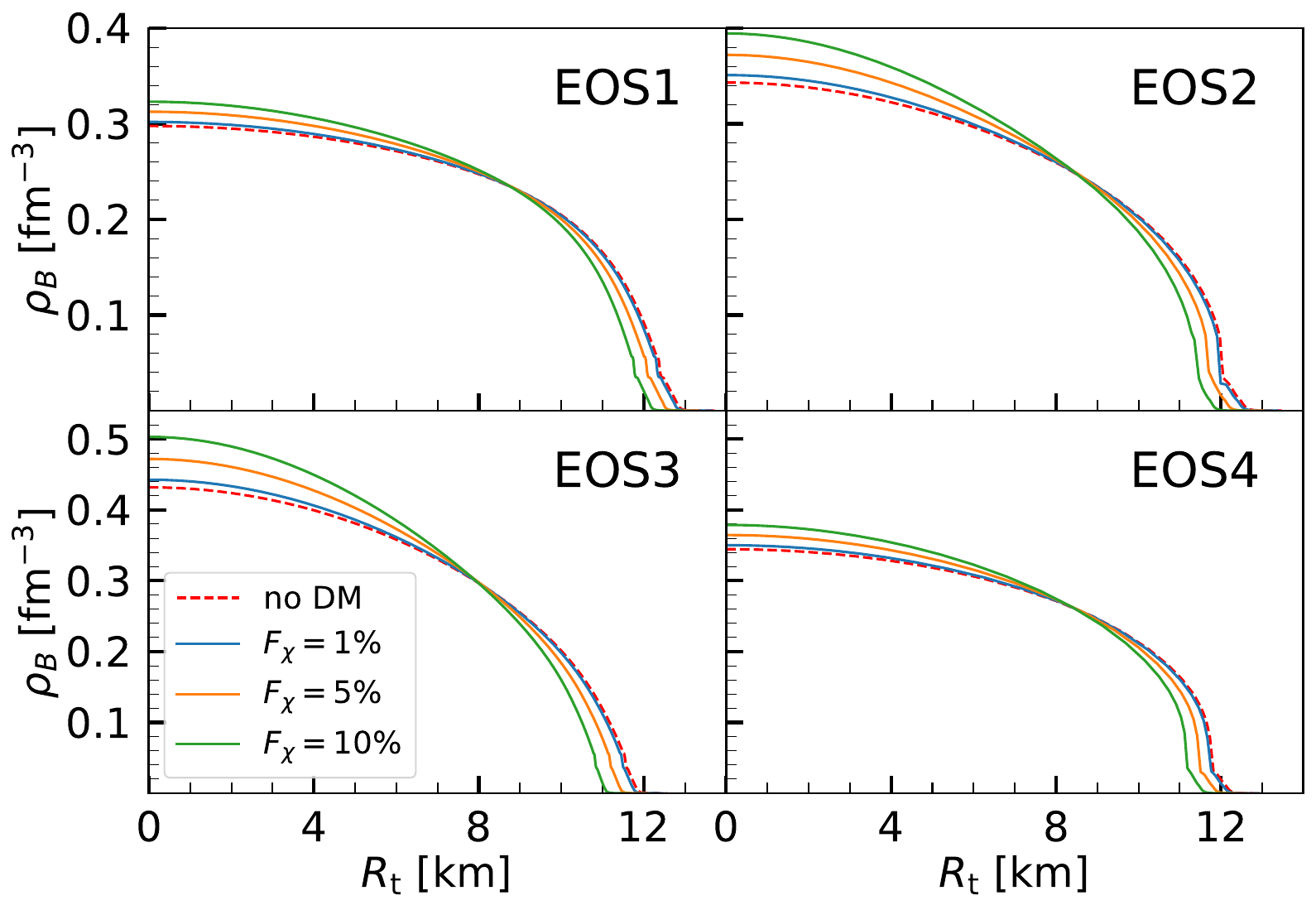}
    \caption{The baryon density $\rho_B$ plotted against the NS radius $R_{\rm t}$ for a 1.4 M$_\odot$ NS. The plot includes four panels, each representing a different nuclear EOS, along with a generic dark matter EOS.}
    \label{fig:nsp}
\end{figure}

\begin{figure}
    \centering
    \includegraphics[width=0.48\textwidth]{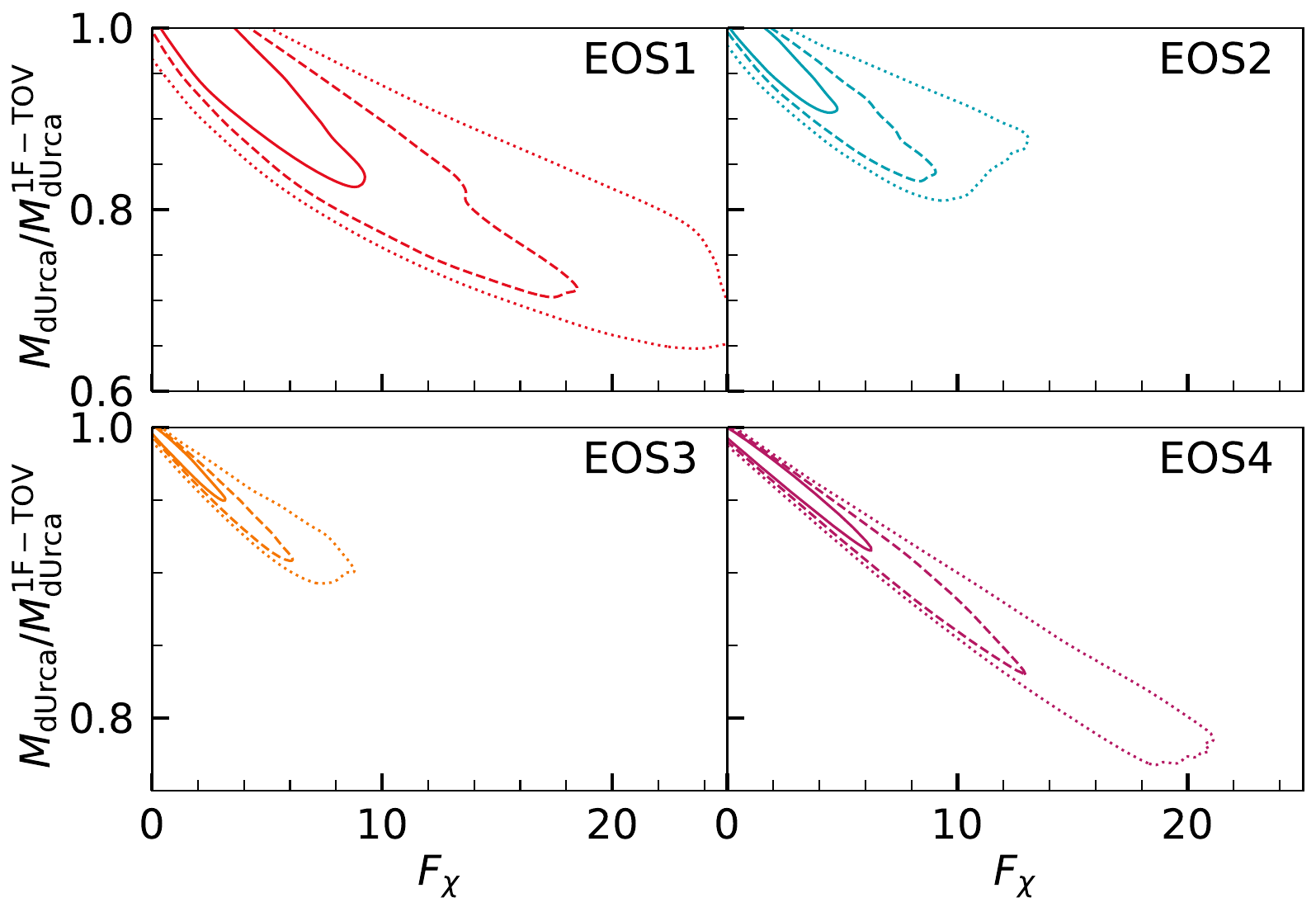}
    \caption{The $M_{dUrca}$/$M_{dUrca}^{1F-TOV}$ as a function of dark matter mass fraction $F_{\chi}$. The top panels are for EOS1 and EOS2, whereas the bottom panel represents results for EOS3 and EOS4. The solid line in each panel represents 68$\%$ confidence interval whereas dashed and dotted lines represent 95$\%$ and 99$\%$ CI, respectively.}
    \label{fig:dUrca_fchi}
\end{figure}

In Figure \ref{fig:rhob_fchi}, we investigate the effect of DM on the NS central density. The main effect is the compression of matter inside a star which results  in a decrease of the NS radius as the fraction of DM increases and the gravitation mass is kept constant. We consider all four equation-of-states (EOSs) to examine how dark matter affects this compression. The plot shows the scaled central density $\rho_B$/$\rho_B^{1F-TOV}$ of stars where $\rho_B^{1F-TOV}$ represents the central baryon density in the absence of dark matter (single fluid TOV) for masses 1.4, 1.6, and 1.8 $M_\odot$ as a function of the percentage of dark matter $F_{\chi}$. The solid line in each panel represents 68$\%$ confidence interval whereas dashed and dotted lines represent 95$\%$ and 99$\%$ CI, respectively. 
From the figure, it is evident that, for each EOS, as the percentage of dark matter ($F_{\chi}$) increases, the central density for masses ranging from 1.4 to 1.8 $M_\odot$ increases in all cases.  In Fig. \ref{fig:nsp}, we plot the density profile of 1.4$M_\odot$ NS with different fractions of DM. The presence of DM increases the gravitational interaction at the star center. As a consequence, mass is pushed to the center,  the central baryonic density  increases, and the radius of the star decreases. This is an interesting result because it indicates that due to the presence of DM, processes that are otherwise not favorable are now allowed, e.g., the onset of hyperons or of the nucleonic direct Urca, etc. may  open in smaller mass stars with the DM presence. The increase in central baryon density due to the compression of matter may also give rise to quark hadron phase transition inside the core of dark matter admixed neutron stars. Model calculations indicate that onset densities for hadron-quark pasta phases and pure quark matter phase can be of the order of 0.4-0.7 fm$^{-3}$~\cite{Ju:2021nev}. These results also imply that the accumulation of dark matter inside neutron stars can trigger the QCD phase transition. This is a novel but model-dependent result and it needs further detailed studies.

In the following, we discuss how the presence of dark matter leads to a decrease in the mass of the star where nucleonic direct Urca processes start to occur, which we designate as $M_{\rm dUrca}$. 
In Figure \ref{fig:dUrca_fchi}, the scaled Urca mass $M_{dUrca}$/$M_{dUrca}^{1F-TOV}$  is plotted as a function of the fraction of dark matter, where  $M_{dUrca}^{1F-TOV}$ represents the Urca mass with single fluid TOV, obtained for the four nuclear matter EOSs. A strong correlation between the dark matter fraction and the Urca mass for each individual case is observed. Notice that the decrease in the $M_{\rm dUrca}$ can be large for stiff EOSs. This is because stiff EOSs can allow a large dark matter fraction inside the neutron stars.

In a previous study conducted by Malik et al. \cite{Malik:2022ilb}, it was argued that the Urca mass exhibits a robust correlation with nuclear symmetry energy. 
The presence of dark matter may, however,  affect our perception of  the central baryonic density, resulting in a wrong estimation of the proton density, in particular, a  larger proton fraction, and, therefore larger nuclear symmetry energy. To gain a comprehensive understanding of these phenomena, further investigations are required in the future.

\begin{figure}
    \centering
    \includegraphics[width=0.5\textwidth]{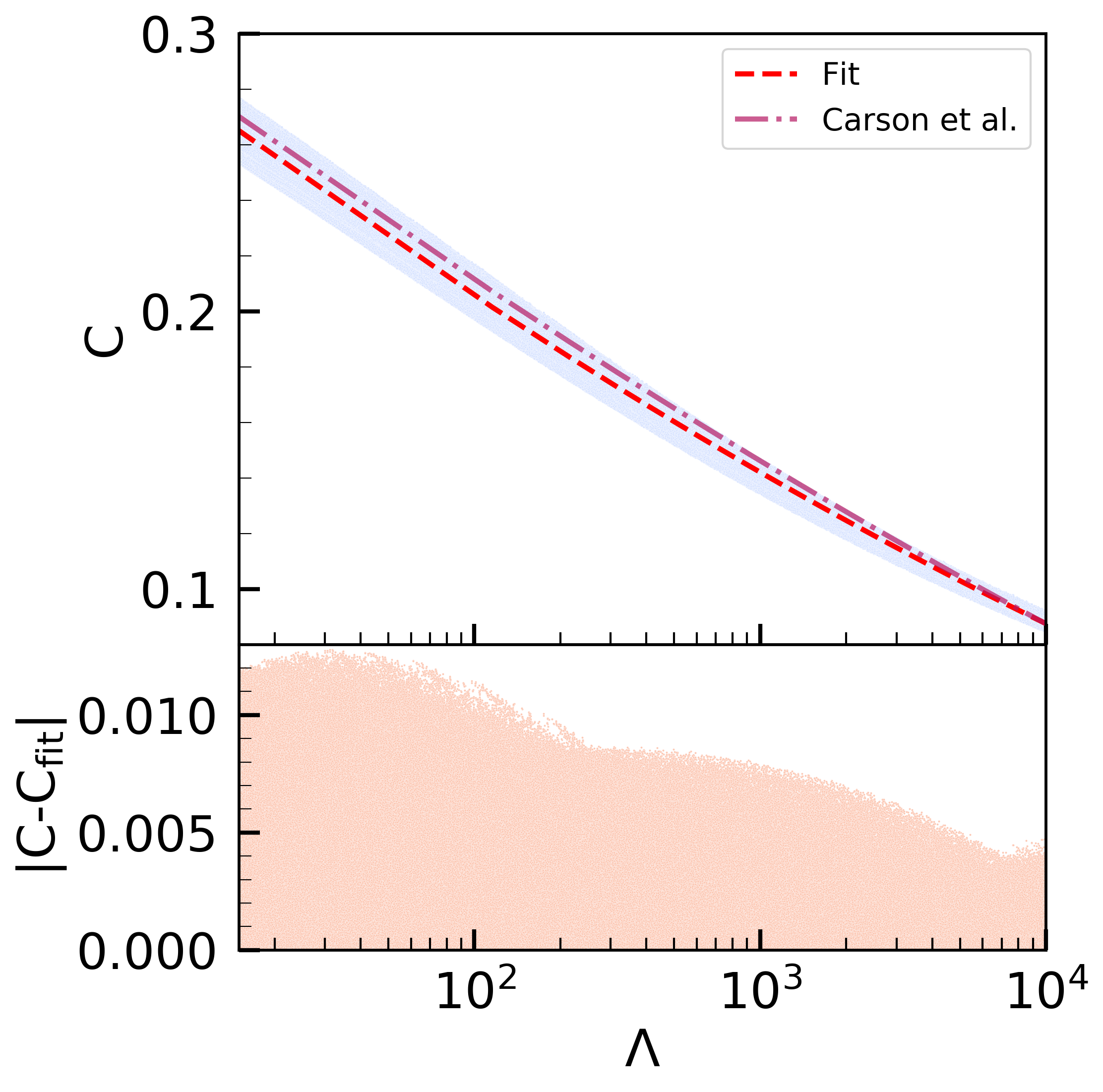}
    \caption{C-Love universal relation for all EOS. The red line is fitted with Eq. \ref{unieq} In the lower panel the residuals for the fitting are calculated. Furthermore, we conduct a comparison with the findings presented in Ref. \cite{Carson:2019rjx} for the scenario of single fluid TOV without dark matter (highlighted in pink-red).}
    \label{fig:universal_1}
\end{figure}

There has been a large interest on the finding of universal relations that involve several NS properties and are independent of the NS mass, see for instance \cite{Yagi:2013bca,Yagi:2016bkt,Maselli:2013mva,Carson:2019rjx}.
Although universal relations for neutron stars are  inherently insensitive to the EOS and, therefore, cannot be utilized to differentiate between different EOS models, they hold the potential to serve as powerful tools for inferring properties of neutron stars in connection with other measurements. Besides, if a particular NS composition breaks the universal relation, this feature may be considered as a smoking gun to identify that special composition. We may, therefore, question whether DM will break some of the known universal relations. 
In the following, we will analyze the universal relation $C-{\Lambda}$ proposed in \cite{Maselli:2013mva}, and discussed in \cite{Carson:2019rjx}. 
The C-Love relationships are depicted in Figure \ref{fig:universal_1}. We begin by fitting the data related to EOS to a simple curve represented by the equation:
\begin{equation}
C = \sum^2_{k=0} a_k (\ln{\Lambda})^k.
\label{unieq}
\end{equation}
By performing this fitting process, we obtain the following values for the coefficients: $a_0 = 0.36054566$, $a_1 = -0.0375908$, and $a_2 = 0.00086283$. In the lower panel of Fig. \ref{fig:universal_1}, the absolute difference from the fits is displayed. The absolute difference is approximately equal to 1$\%$, and furthermore, it decreases to about 0.5$\%$ for higher values of $\Lambda$. We have also compared our results with Figure 4 in Ref. \cite{Carson:2019rjx}, where they used $a_0 = 0.3617$, $a_1 = -0.03548$, and $a_2 = 0.0006194$ for the constrained EOS. This constrained EOS shows an absolute difference of less than 1$\%$, while the unconstrained EOS is around 1$\%$. We conclude that DM does not break the universal relation  $C-\Lambda$.

\section{Conclusions} \label{sec5}
In conclusion, this study has shed light on the connection between dark matter and neutron star properties, while considering the uncertainties in the equation-of-state within the baryonic sector. Dark matter was considered to be formed by fermionic particles with a mass of the order of a few GeVs that interact with a dark vector meson. By employing a two-fluid scenario and sampling 50,000  dark matter EOSs, we analyzed the structure of dark matter admixed neutron stars. It was considered that dark matter is confined within the visible radius of neutron stars, i.e. only no-halo configurations were studied. We have imposed the dark-matter admixed star should have a maximum mass above {1.9}$M_\odot$, a value within $3\sigma$ the PSR J0348+0432 mass, 2.01$\pm 0.04\, M_\odot$. The results revealed interesting correlations between dark matter parameters and various neutron star properties, consistent with results discussed in the literature: the larger the fraction of dark matter the smaller the maximum mass and the smaller the NS radius and tidal deformability \cite{Ellis:2018bkr}. In fact, the dark matter mass fraction within a neutron star was found to have a strong negative correlation with its maximum gravitational mass if a single nuclear model was considered. However, it was shown that this correlation disappears when accounting for the uncertainties associated with nuclear matter EOS.  The maximum mass constraints of dark-matter admixed neutron stars depend on the softness or stiffness of the nuclear matter EOS employed, with some equation-of-states being able to sustain a significant fraction of dark matter, and still describing approximately two solar mass stars.

The inclusion of dark matter led to a decrease in the radius and the  tidal deformability for all masses, indicating the influence of dark matter on the structural characteristics of neutron stars.  Through the analysis of various observational constraints and data, it was demonstrated the potential of dark matter to affect the compression and central energy density of baryonic matter inside neutron stars. The presence of dark-matter originates a nuclear matter compression that translates itself into a larger central baryonic density, a smaller radius and a smaller crust thickness. The increase of the central baryonic density has important consequences on the neutron star properties: it favors the onset of non-nucleonic degrees of freedom  inside less massive stars and may affect the onset of direct Urca processes, affecting the neutron star cooling.  In particular, our study has highlighted the impact of dark matter on the cooling process and nuclear symmetry energy of baryonic matter.
{The detection of stars with similar masses but different surface temperatures could indicate that the cooler ones can be dark-matter admixed stars.}
Additionally, the study explored universal relations, known as the C-Love relationships \cite{Maselli:2013mva,Carson:2019rjx}, which provide insights into neutron star properties that are not easily measurable. We have verified that within 1\% the C-Love universal relation was not broken by the presence of dark-matter, confirming the results of \cite{Ciancarella:2020msu}.

Overall, this research contributes to our understanding of the complex interplay between dark matter and neutron star properties. By uncovering these connections, we are moving closer to unraveling the mysteries concealed within neutron stars.

\begin{table*}[]
\caption{Summary of Neutron Star Properties and their Corresponding Values obtained from Filtered Two-Fluid Solutions with EOS1, EOS2, EOS3, and EOS4. The table presents the median values, along with the lower and upper bounds of the 90\% confidence intervals (CI), for various neutron star properties. These properties include the maximum mass of neutron stars (M$_{\rm max}$), the total radius (R$_{t, x}$) for values of $x$ in the range of [1.2, 1.4, 1.6, 1.8, 2.0], the dimensionless tidal deformability ($\Lambda_{x}$) for $x$ in the range of [1.4, 1.6, 1.8], the fraction of dark matter energy density over nuclear matter ($f_{d,x}$), and the nuclear matter baryon density ($\rho_{B,x}$) for $x$ in the range of [1.4, 1.6, 2.0]. The values are based on comprehensive analyses using EOS1, EOS2, EOS3, and EOS4 as the equation-of-state models for neutron stars. \label{propall}}
\setlength{\tabcolsep}{4.pt}
\renewcommand{\arraystretch}{1.4}
\begin{tabular}{ccccccccccccccccc}
\hline \hline 
\multirow{3}{*}{NS} & \multirow{3}{*}{Units}     & \multicolumn{3}{c}{EOS1}                              &  & \multicolumn{3}{c}{EOS 2}                              &  & \multicolumn{3}{c}{EOS 3}                              &  & \multicolumn{3}{c}{EOS 4}                              \\ \cline{3-5} \cline{7-9} \cline{11-13} \cline{15-17} 
                    &                            & \multirow{2}{*}{median} & \multicolumn{2}{c}{90 \% CI} &  & \multirow{2}{*}{median} & \multicolumn{2}{c}{90 \% CI} &  & \multirow{2}{*}{median} & \multicolumn{2}{c}{90 \% CI} &  & \multirow{2}{*}{median} & \multicolumn{2}{c}{90 \% CI} \\ \cline{4-5} \cline{8-9} \cline{12-13} \cline{16-17} 
                    &                            &                         & min           & max          &  &                         & min           & max          &  &                         & min           & max          &  &                         & min           & max          \\ \hline
M$_{\rm max}$       & $M_\odot$                  & 2.454                   & 2.067         & 2.710        &  & 2.107                   & 2.067         & 2.710        &  & 2.051                   & 2.067         & 2.710        &  & 2.351                   & 2.067         & 2.710        \\
R$_{t,2.0}$         & \multirow{5}{*}{km}        & 13.16                   & 11.77         & 13.87        &  & 12.68                   & 11.77         & 13.87        &  & 11.50                   & 11.77         & 13.87        &  & 12.50                   & 11.77         & 13.87        \\
R$_{t,1.8}$         &                            & 13.16                   & 12.00         & 13.81        &  & 13.00                   & 12.00         & 13.81        &  & 12.04                   & 12.00         & 13.81        &  & 12.56                   & 12.00         & 13.81        \\
R$_{t,1.6}$         &                            & 13.11                   & 12.04         & 13.73        &  & 13.13                   & 12.04         & 13.73        &  & 12.27                   & 12.04         & 13.73        &  & 12.54                   & 12.04         & 13.73        \\
R$_{t,1.4}$         &                            & 13.02                   & 12.01         & 13.63        &  & 13.17                   & 12.01         & 13.63        &  & 12.40                   & 12.01         & 13.63        &  & 12.48                   & 12.01         & 13.63        \\
R$_{t,1.2}$         &                            & 12.92                   & 11.94         & 13.53        &  & 13.18                   & 11.94         & 13.53        &  & 12.48                   & 11.94         & 13.53        &  & 12.40                   & 11.94         & 13.53        \\
$\Lambda_{1.8}$     & \multirow{6}{*}{...}       & 18                      & 7             & 29           &  & 16                      & 7             & 29           &  & 9                       & 7             & 29           &  & 13                      & 7             & 29           \\
$\Lambda_{1.6}$     &                            & 36                      & 17            & 57           &  & 36                      & 17            & 57           &  & 23                      & 17            & 57           &  & 28                      & 17            & 57           \\
$\Lambda_{1.4}$     &                            & 76                      & 39            & 114          &  & 82                      & 39            & 114          &  & 58                      & 39            & 114          &  & 62                      & 39            & 114          \\
$f_{d,2.0}$         &                            & 0.47                    & 0.14          & 0.68         &  & 0.27                    & 0.14          & 0.68         &  & 0.20                    & 0.14          & 0.68         &  & 0.40                    & 0.14          & 0.68         \\
$f_{d,1.6}$         &                            & 0.43                    & 0.12          & 0.65         &  & 0.26                    & 0.12          & 0.65         &  & 0.19                    & 0.12          & 0.65         &  & 0.37                    & 0.12          & 0.65         \\
$f_{d,1.4}$         &                            & 0.41                    & 0.11          & 0.63         &  & 0.24                    & 0.11          & 0.63         &  & 0.18                    & 0.11          & 0.63         &  & 0.34                    & 0.11          & 0.63         \\
$\rho_{B, 2.0}$     & \multirow{3}{*}{fm$^{-3}$} & 0.408                   & 0.350         & 0.576        &  & 0.589                   & 0.350         & 0.576        &  & 0.788                   & 0.350         & 0.576        &  & 0.489                   & 0.350         & 0.576        \\
$\rho_{B, 1.6}$     &                            & 0.339                   & 0.303         & 0.394        &  & 0.399                   & 0.303         & 0.394        &  & 0.500                   & 0.303         & 0.394        &  & 0.395                   & 0.303         & 0.394        \\
$\rho_{B, 1.4}$     &                            & 0.314                   & 0.283         & 0.354        &  & 0.352                   & 0.283         & 0.354        &  & 0.435                   & 0.283         & 0.354        &  & 0.361                   & 0.283         & 0.354        \\ \hline
\end{tabular}
\end{table*}

\section{Acknowledgements} 
The author, P.T, would like to acknowledge CFisUC, University of Coimbra, for their hospitality and local support during his visit in May - June 2023 for the purpose of conducting part of this research. This work was partially supported by national funds from FCT (Fundação para a Ciência e a Tecnologia, I.P, Portugal) under Projects No. UIDP/\-04564/\-2020, No. UIDB/\-04564/\-2020 and 2022.06460.PTDC. T.M. would like to acknowledge the support of FCT – Fundação para a Ciência e a Tecnologia, within Project No. EXPL/FIS-AST/0735/2021. The authors acknowledge the Laboratory for Advanced Computing at the University of Coimbra for providing {HPC} resources that have contributed to the research results reported within this paper, URL: \hyperlink{https://www.uc.pt/lca}{https://www.uc.pt/lca}. A.D. would like to acknowledge Polish National Science Centre Grants No. 2018/30/E/ST2/00432. 

\clearpage
\newpage

\begin{thebibliography}{64}%
\makeatletter
\providecommand \@ifxundefined [1]{%
 \@ifx{#1\undefined}
}%
\providecommand \@ifnum [1]{%
 \ifnum #1\expandafter \@firstoftwo
 \else \expandafter \@secondoftwo
 \fi
}%
\providecommand \@ifx [1]{%
 \ifx #1\expandafter \@firstoftwo
 \else \expandafter \@secondoftwo
 \fi
}%
\providecommand \natexlab [1]{#1}%
\providecommand \enquote  [1]{``#1''}%
\providecommand \bibnamefont  [1]{#1}%
\providecommand \bibfnamefont [1]{#1}%
\providecommand \citenamefont [1]{#1}%
\providecommand \href@noop [0]{\@secondoftwo}%
\providecommand \href [0]{\begingroup \@sanitize@url \@href}%
\providecommand \@href[1]{\@@startlink{#1}\@@href}%
\providecommand \@@href[1]{\endgroup#1\@@endlink}%
\providecommand \@sanitize@url [0]{\catcode `\\12\catcode `\$12\catcode
  `\&12\catcode `\#12\catcode `\^12\catcode `\_12\catcode `\%12\relax}%
\providecommand \@@startlink[1]{}%
\providecommand \@@endlink[0]{}%
\providecommand \url  [0]{\begingroup\@sanitize@url \@url }%
\providecommand \@url [1]{\endgroup\@href {#1}{\urlprefix }}%
\providecommand \urlprefix  [0]{URL }%
\providecommand \Eprint [0]{\href }%
\providecommand \doibase [0]{http://dx.doi.org/}%
\providecommand \selectlanguage [0]{\@gobble}%
\providecommand \bibinfo  [0]{\@secondoftwo}%
\providecommand \bibfield  [0]{\@secondoftwo}%
\providecommand \translation [1]{[#1]}%
\providecommand \BibitemOpen [0]{}%
\providecommand \bibitemStop [0]{}%
\providecommand \bibitemNoStop [0]{.\EOS\space}%
\providecommand \EOS [0]{\spacefactor3000\relax}%
\providecommand \BibitemShut  [1]{\csname bibitem#1\endcsname}%
\let\auto@bib@innerbib\@empty
\bibitem [{\citenamefont {{Glendenning}}(1996)}]{1996cost.book.....G}%
  \BibitemOpen
  \bibfield  {author} {\bibinfo {author} {\bibfnamefont {N.~K.}\ \bibnamefont
  {{Glendenning}}},\ }\href@noop {} {\emph {\bibinfo {title} {{Compact
  Stars}}}}\ (\bibinfo {year} {1996})\BibitemShut {NoStop}%
\bibitem [{\citenamefont {{Haensel}}\ \emph {et~al.}(2007)\citenamefont
  {{Haensel}}, \citenamefont {{Potekhin}},\ and\ \citenamefont
  {{Yakovlev}}}]{book.Haensel2007}%
  \BibitemOpen
  \bibfield  {author} {\bibinfo {author} {\bibfnamefont {P.}~\bibnamefont
  {{Haensel}}}, \bibinfo {author} {\bibfnamefont {A.~Y.}\ \bibnamefont
  {{Potekhin}}}, \ and\ \bibinfo {author} {\bibfnamefont {D.~G.}\ \bibnamefont
  {{Yakovlev}}},\ }\href@noop {} {\emph {\bibinfo {title} {{Neutron Stars 1 :
  Equation of State and Structure}}}},\ Vol.\ \bibinfo {volume} {326}\
  (\bibinfo {year} {2007})\BibitemShut {NoStop}%
\bibitem [{\citenamefont {Rezzolla}\ \emph {et~al.}(2018)\citenamefont
  {Rezzolla}, \citenamefont {Pizzochero}, \citenamefont {Jones}, \citenamefont
  {Rea},\ and\ \citenamefont {Vida\~na}}]{Rezzolla:2018jee}%
  \BibitemOpen
  \bibinfo {editor} {\bibfnamefont {L.}~\bibnamefont {Rezzolla}}, \bibinfo
  {editor} {\bibfnamefont {P.}~\bibnamefont {Pizzochero}}, \bibinfo {editor}
  {\bibfnamefont {D.~I.}\ \bibnamefont {Jones}}, \bibinfo {editor}
  {\bibfnamefont {N.}~\bibnamefont {Rea}}, \ and\ \bibinfo {editor}
  {\bibfnamefont {I.}~\bibnamefont {Vida\~na}},\ eds.,\ \href {\doibase
  10.1007/978-3-319-97616-7} {\emph {\bibinfo {title} {{The Physics and
  Astrophysics of Neutron Stars}}}},\ Vol.\ \bibinfo {volume} {457}\ (\bibinfo
  {publisher} {Springer},\ \bibinfo {year} {2018})\BibitemShut {NoStop}%
\bibitem [{\citenamefont {Lattimer}\ and\ \citenamefont
  {Prakash}(2001)}]{Lattimer:2000nx}%
  \BibitemOpen
  \bibfield  {author} {\bibinfo {author} {\bibfnamefont {J.~M.}\ \bibnamefont
  {Lattimer}}\ and\ \bibinfo {author} {\bibfnamefont {M.}~\bibnamefont
  {Prakash}},\ }\href {\doibase 10.1086/319702} {\bibfield  {journal} {\bibinfo
   {journal} {Astrophys. J.}\ }\textbf {\bibinfo {volume} {550}},\ \bibinfo
  {pages} {426} (\bibinfo {year} {2001})},\ \Eprint
  {http://arxiv.org/abs/astro-ph/0002232} {arXiv:astro-ph/0002232} \BibitemShut
  {NoStop}%
\bibitem [{\citenamefont {Rubin}\ \emph {et~al.}(1978)\citenamefont {Rubin},
  \citenamefont {Ford},\ and\ \citenamefont {Thonnard}}]{Rubin:1978kmz}%
  \BibitemOpen
  \bibfield  {author} {\bibinfo {author} {\bibfnamefont {V.~C.}\ \bibnamefont
  {Rubin}}, \bibinfo {author} {\bibfnamefont {W.~K.}\ \bibnamefont {Ford},
  \bibfnamefont {Jr.}}, \ and\ \bibinfo {author} {\bibfnamefont
  {N.}~\bibnamefont {Thonnard}},\ }\href {\doibase 10.1086/182804} {\bibfield
  {journal} {\bibinfo  {journal} {Astrophys. J. Lett.}\ }\textbf {\bibinfo
  {volume} {225}},\ \bibinfo {pages} {L107} (\bibinfo {year}
  {1978})}\BibitemShut {NoStop}%
\bibitem [{\citenamefont {Bauer}\ and\ \citenamefont
  {Plehn}(2019)}]{Bauer:2017qwy}%
  \BibitemOpen
  \bibfield  {author} {\bibinfo {author} {\bibfnamefont {M.}~\bibnamefont
  {Bauer}}\ and\ \bibinfo {author} {\bibfnamefont {T.}~\bibnamefont {Plehn}},\
  }\href {\doibase 10.1007/978-3-030-16234-4} {\emph {\bibinfo {title} {{Yet
  Another Introduction to Dark Matter}: {The Particle Physics Approach}}}},\
  \bibinfo {series} {Lecture Notes in Physics}, Vol.\ \bibinfo {volume} {959}\
  (\bibinfo  {publisher} {Springer},\ \bibinfo {year} {2019})\ \Eprint
  {http://arxiv.org/abs/1705.01987} {arXiv:1705.01987 [hep-ph]} \BibitemShut
  {NoStop}%
\bibitem [{\citenamefont {Bertone}\ and\ \citenamefont
  {Fairbairn}(2008)}]{Bertone:2007ae}%
  \BibitemOpen
  \bibfield  {author} {\bibinfo {author} {\bibfnamefont {G.}~\bibnamefont
  {Bertone}}\ and\ \bibinfo {author} {\bibfnamefont {M.}~\bibnamefont
  {Fairbairn}},\ }\href {\doibase 10.1103/PhysRevD.77.043515} {\bibfield
  {journal} {\bibinfo  {journal} {Phys. Rev. D}\ }\textbf {\bibinfo {volume}
  {77}},\ \bibinfo {pages} {043515} (\bibinfo {year} {2008})},\ \Eprint
  {http://arxiv.org/abs/0709.1485} {arXiv:0709.1485 [astro-ph]} \BibitemShut
  {NoStop}%
\bibitem [{\citenamefont {de~Lavallaz}\ and\ \citenamefont
  {Fairbairn}(2010)}]{deLavallaz:2010wp}%
  \BibitemOpen
  \bibfield  {author} {\bibinfo {author} {\bibfnamefont {A.}~\bibnamefont
  {de~Lavallaz}}\ and\ \bibinfo {author} {\bibfnamefont {M.}~\bibnamefont
  {Fairbairn}},\ }\href {\doibase 10.1103/PhysRevD.81.123521} {\bibfield
  {journal} {\bibinfo  {journal} {Phys. Rev. D}\ }\textbf {\bibinfo {volume}
  {81}},\ \bibinfo {pages} {123521} (\bibinfo {year} {2010})},\ \Eprint
  {http://arxiv.org/abs/1004.0629} {arXiv:1004.0629 [astro-ph.GA]} \BibitemShut
  {NoStop}%
\bibitem [{\citenamefont {G\"uver}\ \emph {et~al.}(2014)\citenamefont
  {G\"uver}, \citenamefont {Erkoca}, \citenamefont {Hall~Reno},\ and\
  \citenamefont {Sarcevic}}]{Guver:2012ba}%
  \BibitemOpen
  \bibfield  {author} {\bibinfo {author} {\bibfnamefont {T.}~\bibnamefont
  {G\"uver}}, \bibinfo {author} {\bibfnamefont {A.~E.}\ \bibnamefont {Erkoca}},
  \bibinfo {author} {\bibfnamefont {M.}~\bibnamefont {Hall~Reno}}, \ and\
  \bibinfo {author} {\bibfnamefont {I.}~\bibnamefont {Sarcevic}},\ }\href
  {\doibase 10.1088/1475-7516/2014/05/013} {\bibfield  {journal} {\bibinfo
  {journal} {JCAP}\ }\textbf {\bibinfo {volume} {05}},\ \bibinfo {pages} {013}
  (\bibinfo {year} {2014})},\ \Eprint {http://arxiv.org/abs/1201.2400}
  {arXiv:1201.2400 [hep-ph]} \BibitemShut {NoStop}%
\bibitem [{\citenamefont {Goldman}\ and\ \citenamefont
  {Nussinov}(1989)}]{Goldman:1989nd}%
  \BibitemOpen
  \bibfield  {author} {\bibinfo {author} {\bibfnamefont {I.}~\bibnamefont
  {Goldman}}\ and\ \bibinfo {author} {\bibfnamefont {S.}~\bibnamefont
  {Nussinov}},\ }\href {\doibase 10.1103/PhysRevD.40.3221} {\bibfield
  {journal} {\bibinfo  {journal} {Phys. Rev. D}\ }\textbf {\bibinfo {volume}
  {40}},\ \bibinfo {pages} {3221} (\bibinfo {year} {1989})}\BibitemShut
  {NoStop}%
\bibitem [{\citenamefont {Raj}\ \emph {et~al.}(2018)\citenamefont {Raj},
  \citenamefont {Tanedo},\ and\ \citenamefont {Yu}}]{Raj:2017wrv}%
  \BibitemOpen
  \bibfield  {author} {\bibinfo {author} {\bibfnamefont {N.}~\bibnamefont
  {Raj}}, \bibinfo {author} {\bibfnamefont {P.}~\bibnamefont {Tanedo}}, \ and\
  \bibinfo {author} {\bibfnamefont {H.-B.}\ \bibnamefont {Yu}},\ }\href
  {\doibase 10.1103/PhysRevD.97.043006} {\bibfield  {journal} {\bibinfo
  {journal} {Phys. Rev. D}\ }\textbf {\bibinfo {volume} {97}},\ \bibinfo
  {pages} {043006} (\bibinfo {year} {2018})},\ \Eprint
  {http://arxiv.org/abs/1707.09442} {arXiv:1707.09442 [hep-ph]} \BibitemShut
  {NoStop}%
\bibitem [{\citenamefont {Gould}\ \emph {et~al.}(1990)\citenamefont {Gould},
  \citenamefont {Draine}, \citenamefont {Romani},\ and\ \citenamefont
  {Nussinov}}]{Gould:1989gw}%
  \BibitemOpen
  \bibfield  {author} {\bibinfo {author} {\bibfnamefont {A.}~\bibnamefont
  {Gould}}, \bibinfo {author} {\bibfnamefont {B.~T.}\ \bibnamefont {Draine}},
  \bibinfo {author} {\bibfnamefont {R.~W.}\ \bibnamefont {Romani}}, \ and\
  \bibinfo {author} {\bibfnamefont {S.}~\bibnamefont {Nussinov}},\ }\href
  {\doibase 10.1016/0370-2693(90)91745-W} {\bibfield  {journal} {\bibinfo
  {journal} {Phys. Lett. B}\ }\textbf {\bibinfo {volume} {238}},\ \bibinfo
  {pages} {337} (\bibinfo {year} {1990})}\BibitemShut {NoStop}%
\bibitem [{\citenamefont {Kouvaris}(2008)}]{Kouvaris:2007ay}%
  \BibitemOpen
  \bibfield  {author} {\bibinfo {author} {\bibfnamefont {C.}~\bibnamefont
  {Kouvaris}},\ }\href {\doibase 10.1103/PhysRevD.77.023006} {\bibfield
  {journal} {\bibinfo  {journal} {Phys. Rev. D}\ }\textbf {\bibinfo {volume}
  {77}},\ \bibinfo {pages} {023006} (\bibinfo {year} {2008})},\ \Eprint
  {http://arxiv.org/abs/0708.2362} {arXiv:0708.2362 [astro-ph]} \BibitemShut
  {NoStop}%
\bibitem [{\citenamefont {Kouvaris}\ and\ \citenamefont
  {Tinyakov}(2010)}]{Kouvaris:2010vv}%
  \BibitemOpen
  \bibfield  {author} {\bibinfo {author} {\bibfnamefont {C.}~\bibnamefont
  {Kouvaris}}\ and\ \bibinfo {author} {\bibfnamefont {P.}~\bibnamefont
  {Tinyakov}},\ }\href {\doibase 10.1103/PhysRevD.82.063531} {\bibfield
  {journal} {\bibinfo  {journal} {Phys. Rev. D}\ }\textbf {\bibinfo {volume}
  {82}},\ \bibinfo {pages} {063531} (\bibinfo {year} {2010})},\ \Eprint
  {http://arxiv.org/abs/1004.0586} {arXiv:1004.0586 [astro-ph.GA]} \BibitemShut
  {NoStop}%
\bibitem [{\citenamefont {Ellis}\ \emph {et~al.}(2018)\citenamefont {Ellis},
  \citenamefont {H\"utsi}, \citenamefont {Kannike}, \citenamefont {Marzola},
  \citenamefont {Raidal},\ and\ \citenamefont {Vaskonen}}]{Ellis:2018bkr}%
  \BibitemOpen
  \bibfield  {author} {\bibinfo {author} {\bibfnamefont {J.}~\bibnamefont
  {Ellis}}, \bibinfo {author} {\bibfnamefont {G.}~\bibnamefont {H\"utsi}},
  \bibinfo {author} {\bibfnamefont {K.}~\bibnamefont {Kannike}}, \bibinfo
  {author} {\bibfnamefont {L.}~\bibnamefont {Marzola}}, \bibinfo {author}
  {\bibfnamefont {M.}~\bibnamefont {Raidal}}, \ and\ \bibinfo {author}
  {\bibfnamefont {V.}~\bibnamefont {Vaskonen}},\ }\href {\doibase
  10.1103/PhysRevD.97.123007} {\bibfield  {journal} {\bibinfo  {journal} {Phys.
  Rev. D}\ }\textbf {\bibinfo {volume} {97}},\ \bibinfo {pages} {123007}
  (\bibinfo {year} {2018})},\ \Eprint {http://arxiv.org/abs/1804.01418}
  {arXiv:1804.01418 [astro-ph.CO]} \BibitemShut {NoStop}%
\bibitem [{\citenamefont {Panotopoulos}\ and\ \citenamefont
  {Lopes}(2017)}]{Panotopoulos:2017idn}%
  \BibitemOpen
  \bibfield  {author} {\bibinfo {author} {\bibfnamefont {G.}~\bibnamefont
  {Panotopoulos}}\ and\ \bibinfo {author} {\bibfnamefont {I.}~\bibnamefont
  {Lopes}},\ }\href {\doibase 10.1103/PhysRevD.96.083004} {\bibfield  {journal}
  {\bibinfo  {journal} {Phys. Rev. D}\ }\textbf {\bibinfo {volume} {96}},\
  \bibinfo {pages} {083004} (\bibinfo {year} {2017})},\ \Eprint
  {http://arxiv.org/abs/1709.06312} {arXiv:1709.06312 [hep-ph]} \BibitemShut
  {NoStop}%
\bibitem [{\citenamefont {Das}\ \emph {et~al.}(2019)\citenamefont {Das},
  \citenamefont {Malik},\ and\ \citenamefont {Nayak}}]{Das:2018frc}%
  \BibitemOpen
  \bibfield  {author} {\bibinfo {author} {\bibfnamefont {A.}~\bibnamefont
  {Das}}, \bibinfo {author} {\bibfnamefont {T.}~\bibnamefont {Malik}}, \ and\
  \bibinfo {author} {\bibfnamefont {A.~C.}\ \bibnamefont {Nayak}},\ }\href
  {\doibase 10.1103/PhysRevD.99.043016} {\bibfield  {journal} {\bibinfo
  {journal} {Phys. Rev. D}\ }\textbf {\bibinfo {volume} {99}},\ \bibinfo
  {pages} {043016} (\bibinfo {year} {2019})},\ \Eprint
  {http://arxiv.org/abs/1807.10013} {arXiv:1807.10013 [hep-ph]} \BibitemShut
  {NoStop}%
\bibitem [{\citenamefont {Das}\ \emph {et~al.}(2020)\citenamefont {Das},
  \citenamefont {Kumar}, \citenamefont {Kumar}, \citenamefont {Kumar~Biswal},
  \citenamefont {Nakatsukasa}, \citenamefont {Li},\ and\ \citenamefont
  {Patra}}]{Das:2020vng}%
  \BibitemOpen
  \bibfield  {author} {\bibinfo {author} {\bibfnamefont {H.~C.}\ \bibnamefont
  {Das}}, \bibinfo {author} {\bibfnamefont {A.}~\bibnamefont {Kumar}}, \bibinfo
  {author} {\bibfnamefont {B.}~\bibnamefont {Kumar}}, \bibinfo {author}
  {\bibfnamefont {S.}~\bibnamefont {Kumar~Biswal}}, \bibinfo {author}
  {\bibfnamefont {T.}~\bibnamefont {Nakatsukasa}}, \bibinfo {author}
  {\bibfnamefont {A.}~\bibnamefont {Li}}, \ and\ \bibinfo {author}
  {\bibfnamefont {S.~K.}\ \bibnamefont {Patra}},\ }\href {\doibase
  10.1093/mnras/staa1435} {\bibfield  {journal} {\bibinfo  {journal} {Mon. Not.
  Roy. Astron. Soc.}\ }\textbf {\bibinfo {volume} {495}},\ \bibinfo {pages}
  {4893} (\bibinfo {year} {2020})},\ \Eprint {http://arxiv.org/abs/2002.00594}
  {arXiv:2002.00594 [nucl-th]} \BibitemShut {NoStop}%
\bibitem [{\citenamefont {Silk}\ \emph {et~al.}(2010)\citenamefont {Silk} \emph
  {et~al.}}]{Bertone:2010zza}%
  \BibitemOpen
  \bibfield  {author} {\bibinfo {author} {\bibfnamefont {J.}~\bibnamefont
  {Silk}} \emph {et~al.},\ }\href {\doibase 10.1017/CBO9780511770739} {\emph
  {\bibinfo {title} {{Particle Dark Matter: Observations, Models and
  Searches}}}},\ edited by\ \bibinfo {editor} {\bibfnamefont {G.}~\bibnamefont
  {Bertone}}\ (\bibinfo  {publisher} {Cambridge Univ. Press},\ \bibinfo
  {address} {Cambridge},\ \bibinfo {year} {2010})\BibitemShut {NoStop}%
\bibitem [{\citenamefont {Calmet}\ and\ \citenamefont
  {Kuipers}(2021)}]{Calmet:2020pub}%
  \BibitemOpen
  \bibfield  {author} {\bibinfo {author} {\bibfnamefont {X.}~\bibnamefont
  {Calmet}}\ and\ \bibinfo {author} {\bibfnamefont {F.}~\bibnamefont
  {Kuipers}},\ }\href {\doibase 10.1016/j.physletb.2021.136068} {\bibfield
  {journal} {\bibinfo  {journal} {Phys. Lett. B}\ }\textbf {\bibinfo {volume}
  {814}},\ \bibinfo {pages} {136068} (\bibinfo {year} {2021})},\ \Eprint
  {http://arxiv.org/abs/2009.11575} {arXiv:2009.11575 [hep-ph]} \BibitemShut
  {NoStop}%
\bibitem [{\citenamefont {Diedrichs}\ \emph {et~al.}(2023)\citenamefont
  {Diedrichs}, \citenamefont {Becker}, \citenamefont {Jockel}, \citenamefont
  {Christian}, \citenamefont {Sagunski},\ and\ \citenamefont
  {Schaffner-Bielich}}]{Diedrichs:2023trk}%
  \BibitemOpen
  \bibfield  {author} {\bibinfo {author} {\bibfnamefont {R.~F.}\ \bibnamefont
  {Diedrichs}}, \bibinfo {author} {\bibfnamefont {N.}~\bibnamefont {Becker}},
  \bibinfo {author} {\bibfnamefont {C.}~\bibnamefont {Jockel}}, \bibinfo
  {author} {\bibfnamefont {J.-E.}\ \bibnamefont {Christian}}, \bibinfo {author}
  {\bibfnamefont {L.}~\bibnamefont {Sagunski}}, \ and\ \bibinfo {author}
  {\bibfnamefont {J.}~\bibnamefont {Schaffner-Bielich}},\ }\href@noop {} {\
  (\bibinfo {year} {2023})},\ \Eprint {http://arxiv.org/abs/2303.04089}
  {arXiv:2303.04089 [gr-qc]} \BibitemShut {NoStop}%
\bibitem [{\citenamefont {Leung}\ \emph {et~al.}(2022)\citenamefont {Leung},
  \citenamefont {Chu},\ and\ \citenamefont {Lin}}]{Leung:2022wcf}%
  \BibitemOpen
  \bibfield  {author} {\bibinfo {author} {\bibfnamefont {K.-L.}\ \bibnamefont
  {Leung}}, \bibinfo {author} {\bibfnamefont {M.-c.}\ \bibnamefont {Chu}}, \
  and\ \bibinfo {author} {\bibfnamefont {L.-M.}\ \bibnamefont {Lin}},\ }\href
  {\doibase 10.1103/PhysRevD.105.123010} {\bibfield  {journal} {\bibinfo
  {journal} {Phys. Rev. D}\ }\textbf {\bibinfo {volume} {105}},\ \bibinfo
  {pages} {123010} (\bibinfo {year} {2022})},\ \Eprint
  {http://arxiv.org/abs/2207.02433} {arXiv:2207.02433 [astro-ph.HE]}
  \BibitemShut {NoStop}%
\bibitem [{\citenamefont {Abbott}\ \emph {et~al.}(2016)\citenamefont {Abbott},
  \citenamefont {Abbott}, \citenamefont {Abbott}, \citenamefont {Abernathy},
  \citenamefont {Acernese}, \citenamefont {Ackley}, \citenamefont {Adams},
  \citenamefont {Adams}, \citenamefont {Addesso}, \citenamefont {Adhikari}
  \emph {et~al.}}]{abbott2016observation}%
  \BibitemOpen
  \bibfield  {author} {\bibinfo {author} {\bibfnamefont {B.~P.}\ \bibnamefont
  {Abbott}}, \bibinfo {author} {\bibfnamefont {R.}~\bibnamefont {Abbott}},
  \bibinfo {author} {\bibfnamefont {T.}~\bibnamefont {Abbott}}, \bibinfo
  {author} {\bibfnamefont {M.}~\bibnamefont {Abernathy}}, \bibinfo {author}
  {\bibfnamefont {F.}~\bibnamefont {Acernese}}, \bibinfo {author}
  {\bibfnamefont {K.}~\bibnamefont {Ackley}}, \bibinfo {author} {\bibfnamefont
  {C.}~\bibnamefont {Adams}}, \bibinfo {author} {\bibfnamefont
  {T.}~\bibnamefont {Adams}}, \bibinfo {author} {\bibfnamefont
  {P.}~\bibnamefont {Addesso}}, \bibinfo {author} {\bibfnamefont
  {R.}~\bibnamefont {Adhikari}},  \emph {et~al.},\ }\href@noop {} {\bibfield
  {journal} {\bibinfo  {journal} {Physical review letters}\ }\textbf {\bibinfo
  {volume} {116}},\ \bibinfo {pages} {061102} (\bibinfo {year}
  {2016})}\BibitemShut {NoStop}%
\bibitem [{\citenamefont {Abbott}\ \emph {et~al.}(2017)\citenamefont {Abbott},
  \citenamefont {Abbott}, \citenamefont {Abbott}, \citenamefont {Acernese},
  \citenamefont {Ackley}, \citenamefont {Adams}, \citenamefont {Adams},
  \citenamefont {Addesso}, \citenamefont {Adhikari}, \citenamefont {Adya} \emph
  {et~al.}}]{abbott2017gw170817}%
  \BibitemOpen
  \bibfield  {author} {\bibinfo {author} {\bibfnamefont {B.~P.}\ \bibnamefont
  {Abbott}}, \bibinfo {author} {\bibfnamefont {R.}~\bibnamefont {Abbott}},
  \bibinfo {author} {\bibfnamefont {T.}~\bibnamefont {Abbott}}, \bibinfo
  {author} {\bibfnamefont {F.}~\bibnamefont {Acernese}}, \bibinfo {author}
  {\bibfnamefont {K.}~\bibnamefont {Ackley}}, \bibinfo {author} {\bibfnamefont
  {C.}~\bibnamefont {Adams}}, \bibinfo {author} {\bibfnamefont
  {T.}~\bibnamefont {Adams}}, \bibinfo {author} {\bibfnamefont
  {P.}~\bibnamefont {Addesso}}, \bibinfo {author} {\bibfnamefont
  {R.}~\bibnamefont {Adhikari}}, \bibinfo {author} {\bibfnamefont {V.~B.}\
  \bibnamefont {Adya}},  \emph {et~al.},\ }\href@noop {} {\bibfield  {journal}
  {\bibinfo  {journal} {Physical review letters}\ }\textbf {\bibinfo {volume}
  {119}},\ \bibinfo {pages} {161101} (\bibinfo {year} {2017})}\BibitemShut
  {NoStop}%
\bibitem [{\citenamefont {Baiotti}\ and\ \citenamefont
  {Rezzolla}(2017)}]{Baiotti:2016qnr}%
  \BibitemOpen
  \bibfield  {author} {\bibinfo {author} {\bibfnamefont {L.}~\bibnamefont
  {Baiotti}}\ and\ \bibinfo {author} {\bibfnamefont {L.}~\bibnamefont
  {Rezzolla}},\ }\href {\doibase 10.1088/1361-6633/aa67bb} {\bibfield
  {journal} {\bibinfo  {journal} {Rept. Prog. Phys.}\ }\textbf {\bibinfo
  {volume} {80}},\ \bibinfo {pages} {096901} (\bibinfo {year} {2017})},\
  \Eprint {http://arxiv.org/abs/1607.03540} {arXiv:1607.03540 [gr-qc]}
  \BibitemShut {NoStop}%
\bibitem [{\citenamefont {Baiotti}(2019)}]{Baiotti:2019sew}%
  \BibitemOpen
  \bibfield  {author} {\bibinfo {author} {\bibfnamefont {L.}~\bibnamefont
  {Baiotti}},\ }\href {\doibase 10.1016/j.ppnp.2019.103714} {\bibfield
  {journal} {\bibinfo  {journal} {Prog. Part. Nucl. Phys.}\ }\textbf {\bibinfo
  {volume} {109}},\ \bibinfo {pages} {103714} (\bibinfo {year} {2019})},\
  \Eprint {http://arxiv.org/abs/1907.08534} {arXiv:1907.08534 [astro-ph.HE]}
  \BibitemShut {NoStop}%
\bibitem [{\citenamefont {Karkevandi}\ \emph {et~al.}(2022)\citenamefont
  {Karkevandi}, \citenamefont {Shakeri}, \citenamefont {Sagun},\ and\
  \citenamefont {Ivanytskyi}}]{Karkevandi:2021ygv}%
  \BibitemOpen
  \bibfield  {author} {\bibinfo {author} {\bibfnamefont {D.~R.}\ \bibnamefont
  {Karkevandi}}, \bibinfo {author} {\bibfnamefont {S.}~\bibnamefont {Shakeri}},
  \bibinfo {author} {\bibfnamefont {V.}~\bibnamefont {Sagun}}, \ and\ \bibinfo
  {author} {\bibfnamefont {O.}~\bibnamefont {Ivanytskyi}},\ }\href {\doibase
  10.1103/PhysRevD.105.023001} {\bibfield  {journal} {\bibinfo  {journal}
  {Phys. Rev. D}\ }\textbf {\bibinfo {volume} {105}},\ \bibinfo {pages}
  {023001} (\bibinfo {year} {2022})},\ \Eprint
  {http://arxiv.org/abs/2109.03801} {arXiv:2109.03801 [astro-ph.HE]}
  \BibitemShut {NoStop}%
\bibitem [{\citenamefont {Das}\ \emph {et~al.}(2021)\citenamefont {Das},
  \citenamefont {Kumar},\ and\ \citenamefont {Patra}}]{Das:2021yny}%
  \BibitemOpen
  \bibfield  {author} {\bibinfo {author} {\bibfnamefont {H.~C.}\ \bibnamefont
  {Das}}, \bibinfo {author} {\bibfnamefont {A.}~\bibnamefont {Kumar}}, \ and\
  \bibinfo {author} {\bibfnamefont {S.~K.}\ \bibnamefont {Patra}},\ }\href
  {\doibase 10.1103/PhysRevD.104.063028} {\bibfield  {journal} {\bibinfo
  {journal} {Phys. Rev. D}\ }\textbf {\bibinfo {volume} {104}},\ \bibinfo
  {pages} {063028} (\bibinfo {year} {2021})},\ \Eprint
  {http://arxiv.org/abs/2109.01853} {arXiv:2109.01853 [astro-ph.HE]}
  \BibitemShut {NoStop}%
\bibitem [{\citenamefont {Miao}\ \emph {et~al.}(2022)\citenamefont {Miao},
  \citenamefont {Zhu}, \citenamefont {Li},\ and\ \citenamefont
  {Huang}}]{Miao:2022rqj}%
  \BibitemOpen
  \bibfield  {author} {\bibinfo {author} {\bibfnamefont {Z.}~\bibnamefont
  {Miao}}, \bibinfo {author} {\bibfnamefont {Y.}~\bibnamefont {Zhu}}, \bibinfo
  {author} {\bibfnamefont {A.}~\bibnamefont {Li}}, \ and\ \bibinfo {author}
  {\bibfnamefont {F.}~\bibnamefont {Huang}},\ }\href {\doibase
  10.3847/1538-4357/ac8544} {\bibfield  {journal} {\bibinfo  {journal}
  {Astrophys. J.}\ }\textbf {\bibinfo {volume} {936}},\ \bibinfo {pages} {69}
  (\bibinfo {year} {2022})},\ \Eprint {http://arxiv.org/abs/2204.05560}
  {arXiv:2204.05560 [astro-ph.HE]} \BibitemShut {NoStop}%
\bibitem [{\citenamefont {Hippert}\ \emph {et~al.}(2023)\citenamefont
  {Hippert}, \citenamefont {Dillingham}, \citenamefont {Tan}, \citenamefont
  {Curtin}, \citenamefont {Noronha-Hostler},\ and\ \citenamefont
  {Yunes}}]{Hippert:2022snq}%
  \BibitemOpen
  \bibfield  {author} {\bibinfo {author} {\bibfnamefont {M.}~\bibnamefont
  {Hippert}}, \bibinfo {author} {\bibfnamefont {E.}~\bibnamefont {Dillingham}},
  \bibinfo {author} {\bibfnamefont {H.}~\bibnamefont {Tan}}, \bibinfo {author}
  {\bibfnamefont {D.}~\bibnamefont {Curtin}}, \bibinfo {author} {\bibfnamefont
  {J.}~\bibnamefont {Noronha-Hostler}}, \ and\ \bibinfo {author} {\bibfnamefont
  {N.}~\bibnamefont {Yunes}},\ }\href {\doibase 10.1103/PhysRevD.107.115028}
  {\bibfield  {journal} {\bibinfo  {journal} {Phys. Rev. D}\ }\textbf {\bibinfo
  {volume} {107}},\ \bibinfo {pages} {115028} (\bibinfo {year} {2023})},\
  \Eprint {http://arxiv.org/abs/2211.08590} {arXiv:2211.08590 [astro-ph.HE]}
  \BibitemShut {NoStop}%
\bibitem [{\citenamefont {Rutherford}\ \emph {et~al.}(2023)\citenamefont
  {Rutherford}, \citenamefont {Raaijmakers}, \citenamefont
  {Prescod-Weinstein},\ and\ \citenamefont {Watts}}]{Rutherford:2022xeb}%
  \BibitemOpen
  \bibfield  {author} {\bibinfo {author} {\bibfnamefont {N.}~\bibnamefont
  {Rutherford}}, \bibinfo {author} {\bibfnamefont {G.}~\bibnamefont
  {Raaijmakers}}, \bibinfo {author} {\bibfnamefont {C.}~\bibnamefont
  {Prescod-Weinstein}}, \ and\ \bibinfo {author} {\bibfnamefont
  {A.}~\bibnamefont {Watts}},\ }\href {\doibase 10.1103/PhysRevD.107.103051}
  {\bibfield  {journal} {\bibinfo  {journal} {Phys. Rev. D}\ }\textbf {\bibinfo
  {volume} {107}},\ \bibinfo {pages} {103051} (\bibinfo {year} {2023})},\
  \Eprint {http://arxiv.org/abs/2208.03282} {arXiv:2208.03282 [astro-ph.HE]}
  \BibitemShut {NoStop}%
\bibitem [{\citenamefont {Coughlin}\ \emph {et~al.}(2019)\citenamefont
  {Coughlin}, \citenamefont {Dietrich}, \citenamefont {Margalit},\ and\
  \citenamefont {Metzger}}]{coughlin2019multimessenger}%
  \BibitemOpen
  \bibfield  {author} {\bibinfo {author} {\bibfnamefont {M.~W.}\ \bibnamefont
  {Coughlin}}, \bibinfo {author} {\bibfnamefont {T.}~\bibnamefont {Dietrich}},
  \bibinfo {author} {\bibfnamefont {B.}~\bibnamefont {Margalit}}, \ and\
  \bibinfo {author} {\bibfnamefont {B.~D.}\ \bibnamefont {Metzger}},\
  }\href@noop {} {\bibfield  {journal} {\bibinfo  {journal} {Monthly Notices of
  the Royal Astronomical Society: Letters}\ }\textbf {\bibinfo {volume}
  {489}},\ \bibinfo {pages} {L91} (\bibinfo {year} {2019})}\BibitemShut
  {NoStop}%
\bibitem [{\citenamefont {Dietrich}\ \emph {et~al.}(2020)\citenamefont
  {Dietrich}, \citenamefont {Coughlin}, \citenamefont {Pang}, \citenamefont
  {Bulla}, \citenamefont {Heinzel}, \citenamefont {Issa}, \citenamefont
  {Tews},\ and\ \citenamefont {Antier}}]{dietrich2020multimessenger}%
  \BibitemOpen
  \bibfield  {author} {\bibinfo {author} {\bibfnamefont {T.}~\bibnamefont
  {Dietrich}}, \bibinfo {author} {\bibfnamefont {M.~W.}\ \bibnamefont
  {Coughlin}}, \bibinfo {author} {\bibfnamefont {P.~T.}\ \bibnamefont {Pang}},
  \bibinfo {author} {\bibfnamefont {M.}~\bibnamefont {Bulla}}, \bibinfo
  {author} {\bibfnamefont {J.}~\bibnamefont {Heinzel}}, \bibinfo {author}
  {\bibfnamefont {L.}~\bibnamefont {Issa}}, \bibinfo {author} {\bibfnamefont
  {I.}~\bibnamefont {Tews}}, \ and\ \bibinfo {author} {\bibfnamefont
  {S.}~\bibnamefont {Antier}},\ }\href@noop {} {\bibfield  {journal} {\bibinfo
  {journal} {Science}\ }\textbf {\bibinfo {volume} {370}},\ \bibinfo {pages}
  {1450} (\bibinfo {year} {2020})}\BibitemShut {NoStop}%
\bibitem [{\citenamefont {O’Boyle}\ \emph {et~al.}(2020)\citenamefont
  {O’Boyle}, \citenamefont {Markakis}, \citenamefont {Stergioulas},\ and\
  \citenamefont {Read}}]{o2020parametrized}%
  \BibitemOpen
  \bibfield  {author} {\bibinfo {author} {\bibfnamefont {M.~F.}\ \bibnamefont
  {O’Boyle}}, \bibinfo {author} {\bibfnamefont {C.}~\bibnamefont {Markakis}},
  \bibinfo {author} {\bibfnamefont {N.}~\bibnamefont {Stergioulas}}, \ and\
  \bibinfo {author} {\bibfnamefont {J.~S.}\ \bibnamefont {Read}},\ }\href@noop
  {} {\bibfield  {journal} {\bibinfo  {journal} {Physical Review D}\ }\textbf
  {\bibinfo {volume} {102}},\ \bibinfo {pages} {083027} (\bibinfo {year}
  {2020})}\BibitemShut {NoStop}%
\bibitem [{\citenamefont {Pang}\ \emph {et~al.}(2021)\citenamefont {Pang},
  \citenamefont {Tews}, \citenamefont {Coughlin}, \citenamefont {Bulla},
  \citenamefont {Van Den~Broeck},\ and\ \citenamefont
  {Dietrich}}]{pang2021nuclear}%
  \BibitemOpen
  \bibfield  {author} {\bibinfo {author} {\bibfnamefont {P.~T.}\ \bibnamefont
  {Pang}}, \bibinfo {author} {\bibfnamefont {I.}~\bibnamefont {Tews}}, \bibinfo
  {author} {\bibfnamefont {M.~W.}\ \bibnamefont {Coughlin}}, \bibinfo {author}
  {\bibfnamefont {M.}~\bibnamefont {Bulla}}, \bibinfo {author} {\bibfnamefont
  {C.}~\bibnamefont {Van Den~Broeck}}, \ and\ \bibinfo {author} {\bibfnamefont
  {T.}~\bibnamefont {Dietrich}},\ }\href@noop {} {\bibfield  {journal}
  {\bibinfo  {journal} {The Astrophysical Journal}\ }\textbf {\bibinfo {volume}
  {922}},\ \bibinfo {pages} {14} (\bibinfo {year} {2021})}\BibitemShut
  {NoStop}%
\bibitem [{\citenamefont {Pradhan}\ \emph {et~al.}(2022)\citenamefont
  {Pradhan}, \citenamefont {Chatterjee}, \citenamefont {Lanoye},\ and\
  \citenamefont {Jaikumar}}]{Pradhan:2022vdf}%
  \BibitemOpen
  \bibfield  {author} {\bibinfo {author} {\bibfnamefont {B.~K.}\ \bibnamefont
  {Pradhan}}, \bibinfo {author} {\bibfnamefont {D.}~\bibnamefont {Chatterjee}},
  \bibinfo {author} {\bibfnamefont {M.}~\bibnamefont {Lanoye}}, \ and\ \bibinfo
  {author} {\bibfnamefont {P.}~\bibnamefont {Jaikumar}},\ }\href {\doibase
  10.1103/PhysRevC.106.015805} {\bibfield  {journal} {\bibinfo  {journal}
  {Phys. Rev. C}\ }\textbf {\bibinfo {volume} {106}},\ \bibinfo {pages}
  {015805} (\bibinfo {year} {2022})},\ \Eprint
  {http://arxiv.org/abs/2203.03141} {arXiv:2203.03141 [astro-ph.HE]}
  \BibitemShut {NoStop}%
\bibitem [{\citenamefont {Vidana}\ \emph {et~al.}(2009)\citenamefont {Vidana},
  \citenamefont {Provid{\^e}ncia}, \citenamefont {Polls},\ and\ \citenamefont
  {Rios}}]{vidana2009density}%
  \BibitemOpen
  \bibfield  {author} {\bibinfo {author} {\bibfnamefont {I.}~\bibnamefont
  {Vidana}}, \bibinfo {author} {\bibfnamefont {C.}~\bibnamefont
  {Provid{\^e}ncia}}, \bibinfo {author} {\bibfnamefont {A.}~\bibnamefont
  {Polls}}, \ and\ \bibinfo {author} {\bibfnamefont {A.}~\bibnamefont {Rios}},\
  }\href@noop {} {\bibfield  {journal} {\bibinfo  {journal} {Physical Review
  C}\ }\textbf {\bibinfo {volume} {80}},\ \bibinfo {pages} {045806} (\bibinfo
  {year} {2009})}\BibitemShut {NoStop}%
\bibitem [{\citenamefont {Ducoin}\ \emph {et~al.}(2010)\citenamefont {Ducoin},
  \citenamefont {Margueron},\ and\ \citenamefont
  {Providencia}}]{ducoin2010nuclear}%
  \BibitemOpen
  \bibfield  {author} {\bibinfo {author} {\bibfnamefont {C.}~\bibnamefont
  {Ducoin}}, \bibinfo {author} {\bibfnamefont {J.}~\bibnamefont {Margueron}}, \
  and\ \bibinfo {author} {\bibfnamefont {C.}~\bibnamefont {Providencia}},\
  }\href@noop {} {\bibfield  {journal} {\bibinfo  {journal} {EPL (Europhysics
  Letters)}\ }\textbf {\bibinfo {volume} {91}},\ \bibinfo {pages} {32001}
  (\bibinfo {year} {2010})}\BibitemShut {NoStop}%
\bibitem [{\citenamefont {Carson}\ \emph
  {et~al.}(2019{\natexlab{a}})\citenamefont {Carson}, \citenamefont {Steiner},\
  and\ \citenamefont {Yagi}}]{carson2019constraining}%
  \BibitemOpen
  \bibfield  {author} {\bibinfo {author} {\bibfnamefont {Z.}~\bibnamefont
  {Carson}}, \bibinfo {author} {\bibfnamefont {A.~W.}\ \bibnamefont {Steiner}},
  \ and\ \bibinfo {author} {\bibfnamefont {K.}~\bibnamefont {Yagi}},\
  }\href@noop {} {\bibfield  {journal} {\bibinfo  {journal} {Physical Review
  D}\ }\textbf {\bibinfo {volume} {99}},\ \bibinfo {pages} {043010} (\bibinfo
  {year} {2019}{\natexlab{a}})}\BibitemShut {NoStop}%
\bibitem [{\citenamefont {Xie}\ and\ \citenamefont
  {Li}(2019)}]{xie2019bayesian}%
  \BibitemOpen
  \bibfield  {author} {\bibinfo {author} {\bibfnamefont {W.-J.}\ \bibnamefont
  {Xie}}\ and\ \bibinfo {author} {\bibfnamefont {B.-A.}\ \bibnamefont {Li}},\
  }\href@noop {} {\bibfield  {journal} {\bibinfo  {journal} {The Astrophysical
  Journal}\ }\textbf {\bibinfo {volume} {883}},\ \bibinfo {pages} {174}
  (\bibinfo {year} {2019})}\BibitemShut {NoStop}%
\bibitem [{\citenamefont {Zimmerman}\ \emph {et~al.}(2020)\citenamefont
  {Zimmerman}, \citenamefont {Carson}, \citenamefont {Schumacher},
  \citenamefont {Steiner},\ and\ \citenamefont
  {Yagi}}]{zimmerman2020measuring}%
  \BibitemOpen
  \bibfield  {author} {\bibinfo {author} {\bibfnamefont {J.}~\bibnamefont
  {Zimmerman}}, \bibinfo {author} {\bibfnamefont {Z.}~\bibnamefont {Carson}},
  \bibinfo {author} {\bibfnamefont {K.}~\bibnamefont {Schumacher}}, \bibinfo
  {author} {\bibfnamefont {A.~W.}\ \bibnamefont {Steiner}}, \ and\ \bibinfo
  {author} {\bibfnamefont {K.}~\bibnamefont {Yagi}},\ }\href@noop {} {\bibfield
   {journal} {\bibinfo  {journal} {arXiv preprint arXiv:2002.03210}\ }
  (\bibinfo {year} {2020})}\BibitemShut {NoStop}%
\bibitem [{\citenamefont {Das}\ \emph {et~al.}(2022)\citenamefont {Das},
  \citenamefont {Malik},\ and\ \citenamefont {Nayak}}]{Das:2020ecp}%
  \BibitemOpen
  \bibfield  {author} {\bibinfo {author} {\bibfnamefont {A.}~\bibnamefont
  {Das}}, \bibinfo {author} {\bibfnamefont {T.}~\bibnamefont {Malik}}, \ and\
  \bibinfo {author} {\bibfnamefont {A.~C.}\ \bibnamefont {Nayak}},\ }\href
  {\doibase 10.1103/PhysRevD.105.123034} {\bibfield  {journal} {\bibinfo
  {journal} {Phys. Rev. D}\ }\textbf {\bibinfo {volume} {105}},\ \bibinfo
  {pages} {123034} (\bibinfo {year} {2022})},\ \Eprint
  {http://arxiv.org/abs/2011.01318} {arXiv:2011.01318 [nucl-th]} \BibitemShut
  {NoStop}%
\bibitem [{\citenamefont {Xiang}\ \emph {et~al.}(2014)\citenamefont {Xiang},
  \citenamefont {Jiang}, \citenamefont {Zhang},\ and\ \citenamefont
  {Yang}}]{Xiang:2013xwa}%
  \BibitemOpen
  \bibfield  {author} {\bibinfo {author} {\bibfnamefont {Q.-F.}\ \bibnamefont
  {Xiang}}, \bibinfo {author} {\bibfnamefont {W.-Z.}\ \bibnamefont {Jiang}},
  \bibinfo {author} {\bibfnamefont {D.-R.}\ \bibnamefont {Zhang}}, \ and\
  \bibinfo {author} {\bibfnamefont {R.-Y.}\ \bibnamefont {Yang}},\ }\href
  {\doibase 10.1103/PhysRevC.89.025803} {\bibfield  {journal} {\bibinfo
  {journal} {Phys. Rev. C}\ }\textbf {\bibinfo {volume} {89}},\ \bibinfo
  {pages} {025803} (\bibinfo {year} {2014})},\ \Eprint
  {http://arxiv.org/abs/1305.7354} {arXiv:1305.7354 [astro-ph.SR]} \BibitemShut
  {NoStop}%
\bibitem [{\citenamefont {Kouvaris}(2012)}]{kouvaris2012limits}%
  \BibitemOpen
  \bibfield  {author} {\bibinfo {author} {\bibfnamefont {C.}~\bibnamefont
  {Kouvaris}},\ }\href@noop {} {\bibfield  {journal} {\bibinfo  {journal}
  {Physical Review Letters}\ }\textbf {\bibinfo {volume} {108}},\ \bibinfo
  {pages} {191301} (\bibinfo {year} {2012})}\BibitemShut {NoStop}%
\bibitem [{\citenamefont {Bell}\ \emph {et~al.}(2019)\citenamefont {Bell},
  \citenamefont {Busoni},\ and\ \citenamefont {Robles}}]{bell2019capture}%
  \BibitemOpen
  \bibfield  {author} {\bibinfo {author} {\bibfnamefont {N.~F.}\ \bibnamefont
  {Bell}}, \bibinfo {author} {\bibfnamefont {G.}~\bibnamefont {Busoni}}, \ and\
  \bibinfo {author} {\bibfnamefont {S.}~\bibnamefont {Robles}},\ }\href@noop {}
  {\bibfield  {journal} {\bibinfo  {journal} {Journal of Cosmology and
  Astroparticle Physics}\ }\textbf {\bibinfo {volume} {2019}},\ \bibinfo
  {pages} {054} (\bibinfo {year} {2019})}\BibitemShut {NoStop}%
\bibitem [{\citenamefont {Hook}\ and\ \citenamefont
  {Huang}(2018)}]{hook2018probing}%
  \BibitemOpen
  \bibfield  {author} {\bibinfo {author} {\bibfnamefont {A.}~\bibnamefont
  {Hook}}\ and\ \bibinfo {author} {\bibfnamefont {J.}~\bibnamefont {Huang}},\
  }\href@noop {} {\bibfield  {journal} {\bibinfo  {journal} {Journal of High
  Energy Physics}\ }\textbf {\bibinfo {volume} {2018}},\ \bibinfo {pages} {1}
  (\bibinfo {year} {2018})}\BibitemShut {NoStop}%
\bibitem [{\citenamefont {Malik}\ \emph {et~al.}(2023)\citenamefont {Malik},
  \citenamefont {Ferreira}, \citenamefont {Albino},\ and\ \citenamefont
  {Provid\^encia}}]{Malik:2023mnx}%
  \BibitemOpen
  \bibfield  {author} {\bibinfo {author} {\bibfnamefont {T.}~\bibnamefont
  {Malik}}, \bibinfo {author} {\bibfnamefont {M.}~\bibnamefont {Ferreira}},
  \bibinfo {author} {\bibfnamefont {M.~B.}\ \bibnamefont {Albino}}, \ and\
  \bibinfo {author} {\bibfnamefont {C.}~\bibnamefont {Provid\^encia}},\ }\href
  {\doibase 10.1103/PhysRevD.107.103018} {\bibfield  {journal} {\bibinfo
  {journal} {Phys. Rev. D}\ }\textbf {\bibinfo {volume} {107}},\ \bibinfo
  {pages} {103018} (\bibinfo {year} {2023})},\ \Eprint
  {http://arxiv.org/abs/2301.08169} {arXiv:2301.08169 [nucl-th]} \BibitemShut
  {NoStop}%
\bibitem [{\citenamefont {Ciancarella}\ \emph {et~al.}(2021)\citenamefont
  {Ciancarella}, \citenamefont {Pannarale}, \citenamefont {Addazi},\ and\
  \citenamefont {Marciano}}]{Ciancarella:2020msu}%
  \BibitemOpen
  \bibfield  {author} {\bibinfo {author} {\bibfnamefont {R.}~\bibnamefont
  {Ciancarella}}, \bibinfo {author} {\bibfnamefont {F.}~\bibnamefont
  {Pannarale}}, \bibinfo {author} {\bibfnamefont {A.}~\bibnamefont {Addazi}}, \
  and\ \bibinfo {author} {\bibfnamefont {A.}~\bibnamefont {Marciano}},\ }\href
  {\doibase 10.1016/j.dark.2021.100796} {\bibfield  {journal} {\bibinfo
  {journal} {Phys. Dark Univ.}\ }\textbf {\bibinfo {volume} {32}},\ \bibinfo
  {pages} {100796} (\bibinfo {year} {2021})},\ \Eprint
  {http://arxiv.org/abs/2010.12904} {arXiv:2010.12904 [astro-ph.HE]}
  \BibitemShut {NoStop}%
\bibitem [{\citenamefont {Bell}\ \emph {et~al.}(2021)\citenamefont {Bell},
  \citenamefont {Busoni}, \citenamefont {Motta}, \citenamefont {Robles},
  \citenamefont {Thomas},\ and\ \citenamefont {Virgato}}]{Bell:2020obw}%
  \BibitemOpen
  \bibfield  {author} {\bibinfo {author} {\bibfnamefont {N.~F.}\ \bibnamefont
  {Bell}}, \bibinfo {author} {\bibfnamefont {G.}~\bibnamefont {Busoni}},
  \bibinfo {author} {\bibfnamefont {T.~F.}\ \bibnamefont {Motta}}, \bibinfo
  {author} {\bibfnamefont {S.}~\bibnamefont {Robles}}, \bibinfo {author}
  {\bibfnamefont {A.~W.}\ \bibnamefont {Thomas}}, \ and\ \bibinfo {author}
  {\bibfnamefont {M.}~\bibnamefont {Virgato}},\ }\href {\doibase
  10.1103/PhysRevLett.127.111803} {\bibfield  {journal} {\bibinfo  {journal}
  {Phys. Rev. Lett.}\ }\textbf {\bibinfo {volume} {127}},\ \bibinfo {pages}
  {111803} (\bibinfo {year} {2021})},\ \Eprint
  {http://arxiv.org/abs/2012.08918} {arXiv:2012.08918 [hep-ph]} \BibitemShut
  {NoStop}%
\bibitem [{\citenamefont {Jiang}\ \emph {et~al.}(2023)\citenamefont {Jiang},
  \citenamefont {Ecker},\ and\ \citenamefont {Rezzolla}}]{Jiang:2022tps}%
  \BibitemOpen
  \bibfield  {author} {\bibinfo {author} {\bibfnamefont {J.-L.}\ \bibnamefont
  {Jiang}}, \bibinfo {author} {\bibfnamefont {C.}~\bibnamefont {Ecker}}, \ and\
  \bibinfo {author} {\bibfnamefont {L.}~\bibnamefont {Rezzolla}},\ }\href
  {\doibase 10.3847/1538-4357/acc4be} {\bibfield  {journal} {\bibinfo
  {journal} {Astrophys. J.}\ }\textbf {\bibinfo {volume} {949}},\ \bibinfo
  {pages} {11} (\bibinfo {year} {2023})},\ \Eprint
  {http://arxiv.org/abs/2211.00018} {arXiv:2211.00018 [gr-qc]} \BibitemShut
  {NoStop}%
\bibitem [{\citenamefont {Nobleson}\ \emph {et~al.}(2023)\citenamefont
  {Nobleson}, \citenamefont {Banik},\ and\ \citenamefont
  {Malik}}]{Nobleson:2023wca}%
  \BibitemOpen
  \bibfield  {author} {\bibinfo {author} {\bibfnamefont {K.}~\bibnamefont
  {Nobleson}}, \bibinfo {author} {\bibfnamefont {S.}~\bibnamefont {Banik}}, \
  and\ \bibinfo {author} {\bibfnamefont {T.}~\bibnamefont {Malik}},\ }\href
  {\doibase 10.1103/PhysRevD.107.124045} {\bibfield  {journal} {\bibinfo
  {journal} {Phys. Rev. D}\ }\textbf {\bibinfo {volume} {107}},\ \bibinfo
  {pages} {124045} (\bibinfo {year} {2023})},\ \Eprint
  {http://arxiv.org/abs/2306.01054} {arXiv:2306.01054 [gr-qc]} \BibitemShut
  {NoStop}%
\bibitem [{\citenamefont {Abbott}\ \emph {et~al.}(2019)\citenamefont {Abbott}
  \emph {et~al.}}]{LIGOScientific:2018hze}%
  \BibitemOpen
  \bibfield  {author} {\bibinfo {author} {\bibfnamefont {B.~P.}\ \bibnamefont
  {Abbott}} \emph {et~al.} (\bibinfo {collaboration} {LIGO Scientific,
  Virgo}),\ }\href {\doibase 10.1103/PhysRevX.9.011001} {\bibfield  {journal}
  {\bibinfo  {journal} {Phys. Rev. X}\ }\textbf {\bibinfo {volume} {9}},\
  \bibinfo {pages} {011001} (\bibinfo {year} {2019})},\ \Eprint
  {http://arxiv.org/abs/1805.11579} {arXiv:1805.11579 [gr-qc]} \BibitemShut
  {NoStop}%
\bibitem [{\citenamefont {Riley~{\sl et al.}}(2019)}]{Riley:2019yda}%
  \BibitemOpen
  \bibfield  {author} {\bibinfo {author} {\bibfnamefont {T.~E.}\ \bibnamefont
  {Riley~{\sl et al.}}},\ }\href@noop {} {\bibfield  {journal} {\bibinfo
  {journal} {Astrophys. J. Lett.}\ }\textbf {\bibinfo {volume} {887}},\
  \bibinfo {pages} {L21} (\bibinfo {year} {2019})}\BibitemShut {NoStop}%
\bibitem [{\citenamefont {Miller~{\sl et al.}}(2019)}]{Miller:2019cac}%
  \BibitemOpen
  \bibfield  {author} {\bibinfo {author} {\bibfnamefont {M.~C.}\ \bibnamefont
  {Miller~{\sl et al.}}},\ }\href@noop {} {\bibfield  {journal} {\bibinfo
  {journal} {Astrophys. J. Lett.}\ }\textbf {\bibinfo {volume} {887}},\
  \bibinfo {pages} {L24} (\bibinfo {year} {2019})}\BibitemShut {NoStop}%
\bibitem [{\citenamefont {Riley~{\sl et al.}}(2021)}]{Riley:2021pdl}%
  \BibitemOpen
  \bibfield  {author} {\bibinfo {author} {\bibfnamefont {T.~E.}\ \bibnamefont
  {Riley~{\sl et al.}}},\ }\href@noop {} {\bibfield  {journal} {\bibinfo
  {journal} {Astrophys. J. Lett.}\ }\textbf {\bibinfo {volume} {918}},\
  \bibinfo {pages} {L27} (\bibinfo {year} {2021})}\BibitemShut {NoStop}%
\bibitem [{\citenamefont {Miller}\ \emph {et~al.}(2021)\citenamefont {Miller}
  \emph {et~al.}}]{Miller:2021qha}%
  \BibitemOpen
  \bibfield  {author} {\bibinfo {author} {\bibfnamefont {M.~C.}\ \bibnamefont
  {Miller}} \emph {et~al.},\ }\href {\doibase 10.3847/2041-8213/ac089b}
  {\bibfield  {journal} {\bibinfo  {journal} {Astrophys. J. Lett.}\ }\textbf
  {\bibinfo {volume} {918}},\ \bibinfo {pages} {L28} (\bibinfo {year}
  {2021})},\ \Eprint {http://arxiv.org/abs/2105.06979} {arXiv:2105.06979
  [astro-ph.HE]} \BibitemShut {NoStop}%
\bibitem [{\citenamefont {Ivanytskyi}\ \emph {et~al.}(2020)\citenamefont
  {Ivanytskyi}, \citenamefont {Sagun},\ and\ \citenamefont
  {Lopes}}]{Ivanytskyi:2019wxd}%
  \BibitemOpen
  \bibfield  {author} {\bibinfo {author} {\bibfnamefont {O.}~\bibnamefont
  {Ivanytskyi}}, \bibinfo {author} {\bibfnamefont {V.}~\bibnamefont {Sagun}}, \
  and\ \bibinfo {author} {\bibfnamefont {I.}~\bibnamefont {Lopes}},\ }\href
  {\doibase 10.1103/PhysRevD.102.063028} {\bibfield  {journal} {\bibinfo
  {journal} {Phys. Rev. D}\ }\textbf {\bibinfo {volume} {102}},\ \bibinfo
  {pages} {063028} (\bibinfo {year} {2020})},\ \Eprint
  {http://arxiv.org/abs/1910.09925} {arXiv:1910.09925 [astro-ph.HE]}
  \BibitemShut {NoStop}%
\bibitem [{\citenamefont {Louren\c{c}o}\ \emph {et~al.}(2022)\citenamefont
  {Louren\c{c}o}, \citenamefont {Lenzi}, \citenamefont {Frederico},\ and\
  \citenamefont {Dutra}}]{Lourenco:2022fmf}%
  \BibitemOpen
  \bibfield  {author} {\bibinfo {author} {\bibfnamefont {O.}~\bibnamefont
  {Louren\c{c}o}}, \bibinfo {author} {\bibfnamefont {C.~H.}\ \bibnamefont
  {Lenzi}}, \bibinfo {author} {\bibfnamefont {T.}~\bibnamefont {Frederico}}, \
  and\ \bibinfo {author} {\bibfnamefont {M.}~\bibnamefont {Dutra}},\ }\href
  {\doibase 10.1103/PhysRevD.106.043010} {\bibfield  {journal} {\bibinfo
  {journal} {Phys. Rev. D}\ }\textbf {\bibinfo {volume} {106}},\ \bibinfo
  {pages} {043010} (\bibinfo {year} {2022})},\ \Eprint
  {http://arxiv.org/abs/2208.06067} {arXiv:2208.06067 [nucl-th]} \BibitemShut
  {NoStop}%
\bibitem [{\citenamefont {Ju}\ \emph {et~al.}(2021)\citenamefont {Ju},
  \citenamefont {Hu},\ and\ \citenamefont {Shen}}]{Ju:2021nev}%
  \BibitemOpen
  \bibfield  {author} {\bibinfo {author} {\bibfnamefont {M.}~\bibnamefont
  {Ju}}, \bibinfo {author} {\bibfnamefont {J.}~\bibnamefont {Hu}}, \ and\
  \bibinfo {author} {\bibfnamefont {H.}~\bibnamefont {Shen}},\ }\href {\doibase
  10.3847/1538-4357/ac30dd} {\bibfield  {journal} {\bibinfo  {journal}
  {Astrophys. J.}\ }\textbf {\bibinfo {volume} {923}},\ \bibinfo {pages} {250}
  (\bibinfo {year} {2021})},\ \Eprint {http://arxiv.org/abs/2111.08909}
  {arXiv:2111.08909 [nucl-th]} \BibitemShut {NoStop}%
\bibitem [{\citenamefont {Malik}\ \emph {et~al.}(2022)\citenamefont {Malik},
  \citenamefont {Agrawal},\ and\ \citenamefont
  {Provid\^encia}}]{Malik:2022ilb}%
  \BibitemOpen
  \bibfield  {author} {\bibinfo {author} {\bibfnamefont {T.}~\bibnamefont
  {Malik}}, \bibinfo {author} {\bibfnamefont {B.~K.}\ \bibnamefont {Agrawal}},
  \ and\ \bibinfo {author} {\bibfnamefont {C.}~\bibnamefont {Provid\^encia}},\
  }\href {\doibase 10.1103/PhysRevC.106.L042801} {\bibfield  {journal}
  {\bibinfo  {journal} {Phys. Rev. C}\ }\textbf {\bibinfo {volume} {106}},\
  \bibinfo {pages} {L042801} (\bibinfo {year} {2022})},\ \Eprint
  {http://arxiv.org/abs/2206.15404} {arXiv:2206.15404 [nucl-th]} \BibitemShut
  {NoStop}%
\bibitem [{\citenamefont {Carson}\ \emph
  {et~al.}(2019{\natexlab{b}})\citenamefont {Carson}, \citenamefont
  {Chatziioannou}, \citenamefont {Haster}, \citenamefont {Yagi},\ and\
  \citenamefont {Yunes}}]{Carson:2019rjx}%
  \BibitemOpen
  \bibfield  {author} {\bibinfo {author} {\bibfnamefont {Z.}~\bibnamefont
  {Carson}}, \bibinfo {author} {\bibfnamefont {K.}~\bibnamefont
  {Chatziioannou}}, \bibinfo {author} {\bibfnamefont {C.-J.}\ \bibnamefont
  {Haster}}, \bibinfo {author} {\bibfnamefont {K.}~\bibnamefont {Yagi}}, \ and\
  \bibinfo {author} {\bibfnamefont {N.}~\bibnamefont {Yunes}},\ }\href
  {\doibase 10.1103/PhysRevD.99.083016} {\bibfield  {journal} {\bibinfo
  {journal} {Phys. Rev. D}\ }\textbf {\bibinfo {volume} {99}},\ \bibinfo
  {pages} {083016} (\bibinfo {year} {2019}{\natexlab{b}})},\ \Eprint
  {http://arxiv.org/abs/1903.03909} {arXiv:1903.03909 [gr-qc]} \BibitemShut
  {NoStop}%
\bibitem [{\citenamefont {Yagi}\ and\ \citenamefont
  {Yunes}(2013)}]{Yagi:2013bca}%
  \BibitemOpen
  \bibfield  {author} {\bibinfo {author} {\bibfnamefont {K.}~\bibnamefont
  {Yagi}}\ and\ \bibinfo {author} {\bibfnamefont {N.}~\bibnamefont {Yunes}},\
  }\href {\doibase 10.1126/science.1236462} {\bibfield  {journal} {\bibinfo
  {journal} {Science}\ }\textbf {\bibinfo {volume} {341}},\ \bibinfo {pages}
  {365} (\bibinfo {year} {2013})},\ \Eprint {http://arxiv.org/abs/1302.4499}
  {arXiv:1302.4499 [gr-qc]} \BibitemShut {NoStop}%
\bibitem [{\citenamefont {Yagi}\ and\ \citenamefont
  {Yunes}(2017)}]{Yagi:2016bkt}%
  \BibitemOpen
  \bibfield  {author} {\bibinfo {author} {\bibfnamefont {K.}~\bibnamefont
  {Yagi}}\ and\ \bibinfo {author} {\bibfnamefont {N.}~\bibnamefont {Yunes}},\
  }\href {\doibase 10.1016/j.physrep.2017.03.002} {\bibfield  {journal}
  {\bibinfo  {journal} {Phys. Rept.}\ }\textbf {\bibinfo {volume} {681}},\
  \bibinfo {pages} {1} (\bibinfo {year} {2017})},\ \Eprint
  {http://arxiv.org/abs/1608.02582} {arXiv:1608.02582 [gr-qc]} \BibitemShut
  {NoStop}%
\bibitem [{\citenamefont {Maselli}\ \emph {et~al.}(2013)\citenamefont
  {Maselli}, \citenamefont {Cardoso}, \citenamefont {Ferrari}, \citenamefont
  {Gualtieri},\ and\ \citenamefont {Pani}}]{Maselli:2013mva}%
  \BibitemOpen
  \bibfield  {author} {\bibinfo {author} {\bibfnamefont {A.}~\bibnamefont
  {Maselli}}, \bibinfo {author} {\bibfnamefont {V.}~\bibnamefont {Cardoso}},
  \bibinfo {author} {\bibfnamefont {V.}~\bibnamefont {Ferrari}}, \bibinfo
  {author} {\bibfnamefont {L.}~\bibnamefont {Gualtieri}}, \ and\ \bibinfo
  {author} {\bibfnamefont {P.}~\bibnamefont {Pani}},\ }\href {\doibase
  10.1103/PhysRevD.88.023007} {\bibfield  {journal} {\bibinfo  {journal} {Phys.
  Rev. D}\ }\textbf {\bibinfo {volume} {88}},\ \bibinfo {pages} {023007}
  (\bibinfo {year} {2013})},\ \Eprint {http://arxiv.org/abs/1304.2052}
  {arXiv:1304.2052 [gr-qc]} \BibitemShut {NoStop}%
\end{thebibliography}
%
\end{document}